# Impacts of surface chemistry and adsorbed ions on dynamics of water around detonation nanodiamond in aqueous salt solutions

Farshad Saberi-Movahed* and Donald W Brenner

Department of Materials Science and Engineering, North Carolina State University, Raleigh, NC, USA.
(*Corresponding author: fsaberi@ncsu.edu)

## ABSTRACT

Water near detonation nanodiamonds (DNDs) forms a Hydrogen Bond (HB) network, whose strength influences DNDs' fluorescence intensity and colloidal stability in aqueous suspensions. However, effects of dissolved ions and DND's surface chemistry on dynamics of water that manifest in rupture and formation of HBs still remain to be elucidated. Thus, we carried out molecular dynamics simulations to investigate the aforementioned effects in the aqueous salt (any of KCl, NaCl, $CaCl_2$, or $MgCl_2$) solution of DND functionalized with any of –H, $–NH_2$, –COOH, or –OH groups. We observed the specific cation effects on both translational and reorientational dynamics of water around the negatively charged DND–COOH. In the whole hydration shell of this DND, we obtained $K^+ < Na^+ < Ca^{2+} < Mg^{2+}$ ordering for the impact of the cation on reducing the translational diffusion coefficient of water. In the immediate vicinity of the charged DND–COOH, the slowdown impacts of cations on the reorientational dynamics of dipole and OH vectors of water were according to $Na^+ < K^+ < Ca^{2+} < Mg^{2+}$ and $Na^+ < Ca^{2+} < K^+ < Mg^{2+}$ ordering, respectively. Furthermore, regardless of the type of dissolved ions, positively charged DND–$NH_2$ and negatively charged DND–COOH induced, respectively, the slowest dipole and OH reorientational dynamics in the first hydration layer of DND. Our results led us to conclude that charged groups on the surface of DNDs on the one hand and the adsorbed counterions on the other hand cooperatively slow down the reorientational dynamics of water in multiple directions.

**Keywords:** Nanodiamond; Functional Group; Ion; Aqueous Solution; Self-diffusion; Reorientational Dynamics; Wobbling-in-cone Motion; Molecular Dynamics Simulation.

## 1 Introduction

Many of exceptional properties of water, such as the unusual density of liquid water at 4 °C and relatively high surface tension, have been attributed to its hydrogen bond (HB) network[1]. This network has a dynamic nature, which results from the breaking and making of HB bonds in the network due to thermal excitations. These network reconstructions are not possible without rotations and translations of the network's constituent water molecules.[2,3] Therefore, both translational and rotational dynamics of water have been the subject of numerous studies of both experimental and theoretical nature.[4–12] In particular, specific attentions have been paid to the dynamics of the interfacial water around solutes.[13–16]

The motivation behind studying the dynamics of the interfacial water arises from understanding the mutual effects of solutes and water on each other. On the one hand, the functioning of certain molecules such as proteins depends on the HB network of water and its fluctuations and dynamic rearrangements around them.[17–23] For instance, the successful catalytic activity of enzymes requires that the corresponding proteins undergo conformational transitions as smoothly as possible. It has been suggested that the dynamic rearrangements of the interfacial water facilitate these conformational transitions.[24] On the other hand, some solutes such as ions can substantially modify the arrangement of their immediate surrounding water. This phenomenon can have different manifestations in a multitude of macroscopic properties of electrolyte solutions. These effects of ions, which ultimately depend on their size and electric charge, have been referred to as “specific ion effects”.

The “specific ion effects”, was first discovered by Hofmeister in the study of protein solubility in salt solutions.[25] Initially, Hofmeister ordered constituent ions of salts into the so-called Hofmeister series according to their effects on the solubility and conformational stability of proteins in electrolyte solutions.[26] Over time, the series has been expanded to include more salts and also to explain wider range of ionic solution properties such as diffusion coefficient and viscosity.[27,28]

Ions in the Hofmeister series have been further classified into kosmotropes (“structure-makers”) or chaotropes (“structure-breakers”).[29] The former, which include small ions with high surface charge densities such as $Ca^{2+}$ or $Mg^{2+}$, induce highly ordered arrangements in their tightly bound hydration shells. However, the latter, which have larger size and smaller charge densities such as $K^{+}$ or $Cl^{-}$ ions, are characterized by disorganized water structure and weakened HB in their surroundings.[30,31] Recent studies have shown that these structural perturbations are limited to the first hydration shell of most ions, except in certain conditions. More specifically, if both cation and anion of a salt possess high surface charge densities, they can cooperatively slow down the reorientational dynamics of water well beyond their first hydration shells.[32–34] This thereby can lead to locking in the HB network of water in multiple directions.[35]

To the best of our knowledge, the dynamics of hydration shells of Detonation Nanodiamond (DND) in electrolyte solutions have not been studied yet, despite their promising biomedical applications such as anti-cancer drug delivery[36–39], treatments for neurodegenerative diseases[40], and bioimaging[41]. In these applications, DNDs are inevitably in contact with water, certain ions, and some other residues[42–45]. In particular, it has been shown that the HB network of water plays an important role in fluorescence properties of DNDs.[46]

We therefore aimed at investigating the dynamics of water in the hydration shells of DNDs in aqueous solution of inorganic salts. The center of our attention is placed on the effects of the surface chemistry of DNDs, on the one hand, and the solvated salts, on the other hand, on the translational and orientational dynamics of the interfacial water. Particularly, our main objective is to find out the underlying mechanism behind the reorientation of water in the hydration shells, which has paramount implications for HB network reconstructions.

In order to study the translational dynamics of the interfacial water, we have calculated its self-diffusion coefficient using the well-known Einstein relation[47]. Then, we have developed a multiple regression model, which describes the relationship between the self-diffusion coefficient of the interfacial water, on the one hand, and the DNDs’ surface chemistry and solvated salts on the other hand.

To study the rotational dynamics of the interfacial water, we have defined reorientational correlation functions (RCF) to track the time evolution of water’s dipole moment and OH bond reorientations. In a sense, these functions, which have been extensively used in the literature[48], demonstrate how quickly water loses the memory of its original orientations[3]. As pointed out in many studies[4,19,32,49–54], there are multiple timescales in the decay behavior of RCFs. We have used the extended wobbling-in-a-cone (EWIC) model in order to interpret these multiscale reorientational relaxations.[10,52,55,56] This has eventually led us to propose a cooperative mechanism for the rotational dynamics of water in hydration shells of charged DNDs surrounded by counterions. We have presented the details of this mechanism in the remainder of this paper.

We have utilized the classical Molecular Dynamics (MD) simulation[57,58] to carry out the studies explained above. It serves as a great tool to realize our goals for two reasons. First, we can readily use the resulting atomic trajectories to calculate the Mean Squared Displacements (MSD) of water oxygen atoms, which eventually gives us the self-diffusion coefficient of the interfacial water using the Einstein relation. Second, it has been shown that MD simulations have this benefit over most experimental techniques that can capture all three timescales of the reorientational relaxations of the interfacial water.[3,10,52] In addition, the explicit representation of water in the fully atomistic MD simulation enables us to readily identify water orientations along its dipole moment and OH bonds.

The rest of the present paper is organized as follows. We have described the MD simulation setup as well as the utilized computational tools in Section 2. The self-diffusion coefficient of the interfacial water as well as the corresponding multiple regression model are presented in Section 3. Then, we have demonstrated the RCF plots in Section 4 and have discussed the reorientational behavior of water in hydration shells of DNDs and ions. The cooperative mechanism for water's rotational dynamics nearby charged DNDs is also discussed. Finally, we have summarized our important findings in Section 5.

# 2 Methodology

## 2.1 MD simulation setup

Each atomistic system in our MD simulations consists of a single cuboctahedral DND of 4.4 nm in diameter with a distinct surface chemistry, solvated in 0.1M aqueous solution of any of KCl, NaCl, $CaCl_2$, or $MgCl_2$ salts. The charge density of ions making up these salts varies in the order $Cl^- < K^+ < Na^+ < Ca^{2+} < Mg^{2+}$. Four types of surface functionalizations for DNDs are considered, namely, fully hydrogenated (DND–H), hydroxylated (DND–OH), carboxylated (DND–COOH), and aminated (DND–$NH_2$). The last three types of DNDs also have some amounts of hydrogen on their surfaces. Furthermore, DND–OH exists as an electrically neutral particle in the present work, while all other DNDs have both neutral and charged variants. The charged versions of DND–COOH and DND–$NH_2$ are created by adding –$COO^-$ and –$NH_3^+$ to their respective facets, while the charged DND–H is obtained via a charge equilibration method[59]. The detailed, step-by-step procedure that we took to run the MD simulations is described elsewhere[60].

## 2.2 Computational Tools

In this section, we introduce some computational tools that we have developed as adds-on to MDAnalysis[61,62] package. We have used them to post-process the atomic trajectories for characterizing dynamics of water in hydration layers around solutes. The tools are implemented in python codes, which incorporate the Message Passing Interface (MPI) for parallel computing to distribute heavy computations on massive atomic data across high-performance computing clusters.

### 2.2.1 Self-diffusion Coefficient

The self-diffusion coefficient is often employed in MD simulations to characterize the translational motions of atomic systems. Furthermore, it is viable to experimentally measure the self-diffusion coefficient using various Nuclear Magnetic Resonance (NMR) techniques[63,64], hence it provides a validation tool for MD results. In the present study, we are particularly interested in quantifying the effects of DND's surface functional groups and charge density, on the one hand, and dissolved ions, on the other hand, on the translational mobility of water molecules.

The self-diffusion coefficient $D$ of atoms in a 3D system can be obtained from the Einstein relation[47] as

$$D = \lim_{t\to\infty} \frac{\langle[\vec{r}(t) - \vec{r}(t_0)]^2\rangle}{6t} \quad \text{Eq. 1}$$

where the numerator is the MSD, $\vec{r}(t)$ is the coordinate vector of the atom at time $t$, $t_0$ denotes different starting times, and $\langle\cdots\rangle$ is the ensemble average. The Einstein relation tells us that if the MSD vs. time curve turns out to be linear, then its slope divided by 6 gives the self-diffusion coefficient.

### 2.2.2 Reorientational Correlation Function

To obtain a more complete picture of effects of ions and DNDs on the dynamics of water in DNDs' hydration layers, we also need to study the rotational dynamics of water. For this purpose, we use the Reorientational Correlation Function (RCF)[48] defined as

$$C_2^u(t) = \frac{\langle P_2[\vec{u}(t).\vec{u}(t_0)]\rangle}{\langle P_2[\vec{u}(t_0).\vec{u}(t_0)]\rangle} \quad \text{Eq. 2}$$

where $P_2$ is the Legendre polynomial of order 2, $\vec{u}(t)$ is the unit vector along the orientation of interest $\vec{u}$, $t_0$ represents different starting points in time, and $\langle\cdots\rangle$ denotes the ensemble average. In this study, we have investigated RCFs for water's orientations along its dipole moment ($\vec{\mu}$) and OH bonds (see Figure 1(a)). Hereafter, we denote their corresponding correlation functions and related parameters with superscripts *dip* and *oh* (e.g., $C_2^{dip}(t)$ and $C_2^{oh}(t)$), respectively. It is worth noting that $C_2^u(t)$ can be measured via various experimental techniques, of which an in-depth review can be found elsewhere[3].

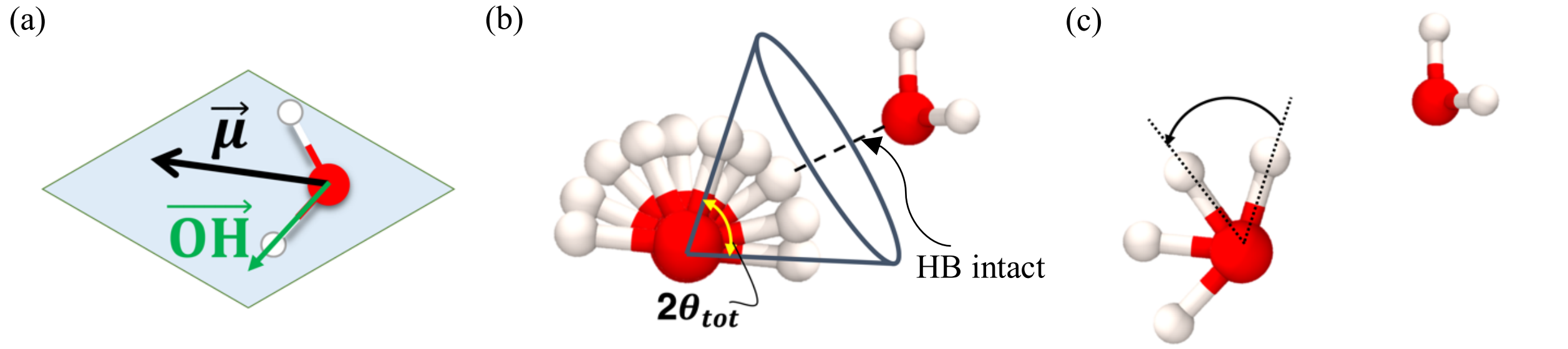

**Figure 1.** (a) Two different orientations of water, that is, dipole moment ($\vec{\mu}$) and $\overrightarrow{OH}$ bond, (b) the hypothetical cone in which $\overrightarrow{OH}$ bond of water carries out inertial and wobbling diffusive motions, both of which result from restricted rotation due to the intact HB with the nearby water, (c) unrestricted rotation of the water molecule that has broken the HB with its neighbor.

In spite of the simple mathematical expression for $C_2^u(t)$, its decay behavior involves complex mechanisms that contribute to the reorientational dynamics of water. Indeed, numerous experimental and theoretical studies have shown that water's reorientational relaxation in confined environments takes place at three different timescales, independent of each other.[6] To provide a concrete interpretation for the multiscale decay of RCFs, we have adopted the extended wobbling-in-a-cone (EWIC) model[10,52,55]. The EWIC model fits a tri-exponential function to $C_2^u(t)$, which is written as

$$C_2^u(t) = \underbrace{\left[(T^u)^2 + (1-(T^u)^2)\exp\left(-\frac{t}{\tau_{in}^u}\right)\right]}_{\text{Inertial motion in a cone with semi-angle } \theta_{in}^u} \times \underbrace{\left[(S^u)^2 + (1-(S^u)^2)\exp\left(-\frac{t}{\tau_c^u}\right)\right]}_{\text{Wobbling motion in a cone with semi-angle } \theta_c^u} \times \underbrace{\exp\left(-\frac{t}{\tau_m^u}\right)}_{\text{Final unrestricted diffusive motion}} \qquad \text{Eq. 3}$$

According to the EWIC model, water molecules first undergo an inertial-liberational motion in a cone with the semi-angle of $\theta_{in}^u$, followed by wobbling motions in a cone with a larger semi-angle $\theta_c^u$ than $\theta_{in}^u$ (see Figure 1(b)). Finally, the rotational diffusion of water without angular restrictions leads to its complete orientational randomization (see Figure 1(c)). In a sense, these three rotational modes of water have been reflected in the functional form of the EWIC model in Eq. 3. That is, the inertial reorientation decays with the relaxation time $\tau_{in}^u$ to a plateau, which is further decayed by the wobbling diffusive motion in the cone to a second plateau with the relaxation time constant of $\tau_c^u$. The second plateau is then further decayed by the final rotational diffusion process with the relaxation time $\tau_m^u$. While the relaxation of the liberational reorientation and the wobbling diffusion occur, respectively, on sub-ps and ~1- ps timescales, the final rotational diffusion process has the relaxation timescales of 10s-100s ps.

Tan *et al.* have related the angularly restricted (i.e., liberational and wobbling) and unrestricted (rotational diffusion) modes of water reorientations to the dynamics of water's HB network.[52] While the former reflects the local fluctuations in the HB network without breaking the HBs, the latter leads to the breaking and making of HBs and hence is thought of as a global process. According to the EWIC model, we can quantify these local and global processes by using, respectively, the wobbling diffusion coefficient $D_c^u$ and the rotational diffusion coefficient $D_m^u$, which are defined as:

$$D_c^u = \frac{(x_c^u)^2(1+x_c^u)^2\{\ln[(1+x_c^u)/2] + (1-x_c^u)/2\}}{\tau_c^u(1-(S^u)^2)[2(x_c^u-1)]} + \frac{(1-x_c^u)(6+8x_c^u-(x_c^u)^2-12(x_c^u)^3-7(x_c^u)^4)}{24\tau_c^u(1-(S^u)^2)} \qquad \text{Eq. 4}$$

$$(S^u)^2 = \left[\frac{1}{2}(\cos\theta_c^u)(1+\cos\theta_c^u)\right]^2 \qquad \text{Eq. 5}$$

$$D_m^u = \frac{1}{6\tau_m^u} \qquad \text{Eq. 6}$$

where $x_c^u = \cos\theta_c^u$. We can replace $(S^u)^2$ by $(T^u)^2(S^u)^2$ in Eq. 5 to obtain the total cone angle $\theta_{tot}^u$, which denotes angles sampled by both inertial and wobbling diffusive motions (see Figure 1(b)). Furthermore, we calculate the orientational correlation time[65], $\tau_{corr}^u$, as the time integration of $C_2^u(t)$

$$\tau_{corr}^u = \int_0^\infty C_2^u(t)dt \qquad \text{Eq. 7}$$

For the completeness of the discussion, we need to briefly mention a different competing model, called the Extended Jump Model (EJM)[66,67], which has recently attracted some attentions[19,68]. This model involves two unequally weighted contributions to the reorientation of water in its local environment: 1) dominant contributions from the exchange of HB acceptors, which leads to sudden large amplitude jumps of water's OH bonds, 2) angular diffusive rotations of intact HBs in between the jumps. Although the first mode of EJM is absent in the EWIC model, the second contribution to water reorientations resembles to the diffusive wobbling motion of water's orientations in the hypothetical cone. However, EJM only focuses on the local changes in the orientation of water and does not explain the final rotational diffusion process, which leads to the global reconstructions in the HB network[3]. Thus, we decided to adopt the EWIC in our study.

To calculate RCFs, we have sampled conformations during MD simulations every 100 steps from the last 200 ps of the NVT production run. It means that the sampling time window is 0.1 ps given that every MD step is 1 fs. The rationale behind this choice is twofold. First, it helps capture the fast, sub-picosecond mode of the reorientational dynamics to some acceptable extent. Secondly, if we had chosen smaller frequencies to save atomic trajectories, the computational task would have been prohibitively difficult.

## 3 Results and Discussions

### 3.1 Translational dynamics

We study the translational motion of water around DNDs by measuring water's self-diffusion coefficient *D*. The Einstein relation (see Eq. 1) gives *D* as the slope of the MSD vs. time, provided that they are linearly related. We calculated MSDs of water's oxygen atom in the whole hydration shell of various DNDs solvated in different salt solutions. The whole hydration shell is defined as a 1 nm thick spherical shell around the DND and centered at its centroid, beyond which water has a bulk-like behavior. Roughly speaking, it is an aggregation of the first to the third hydration layers of the DND, which we had identified in our previous study[60]. Our initial goal was to calculate the self-diffusion coefficient of water in each of the first, the second, and the third hydration layers of DNDs. However, the relatively small number of water molecules in those layers makes the MSD of water suffer from a lack of sufficient statistical sampling. Thus, we instead decided to calculate the self-diffusion coefficient of water in the whole hydration layer.

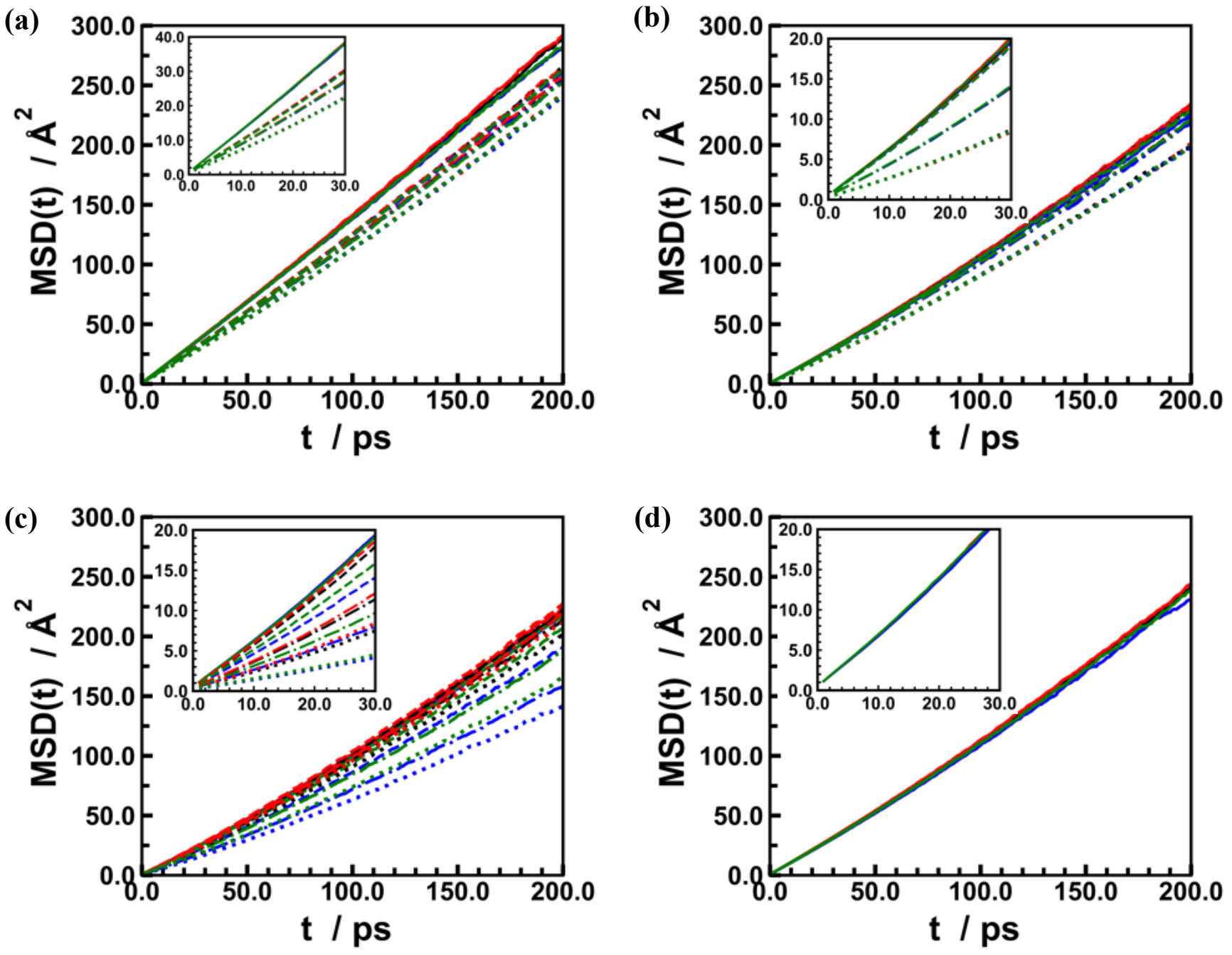

**Figure 2.** MSDs of water in the whole hydration layer of different DNDs with various surface charges that are solvated in four different salt solutions: KCl (red line), NaCl (black line), $CaCl_2$ (green line), $MgCl_2$ (blue line). (a) DND–H, (b) DND–$NH_2$, (c) DND–COOH, (d) DND–OH. Solid, dashed, dash-dotted, and dotted lines correspond to 0, 28, 56, and 84 absolute charges on DNDs, respectively. DND–H and DND–$NH_2$ assume any of these charges with the positive sign, so does DND–COOH but with the negative sign. But DND–OH only exists as a neutral particle in our study.

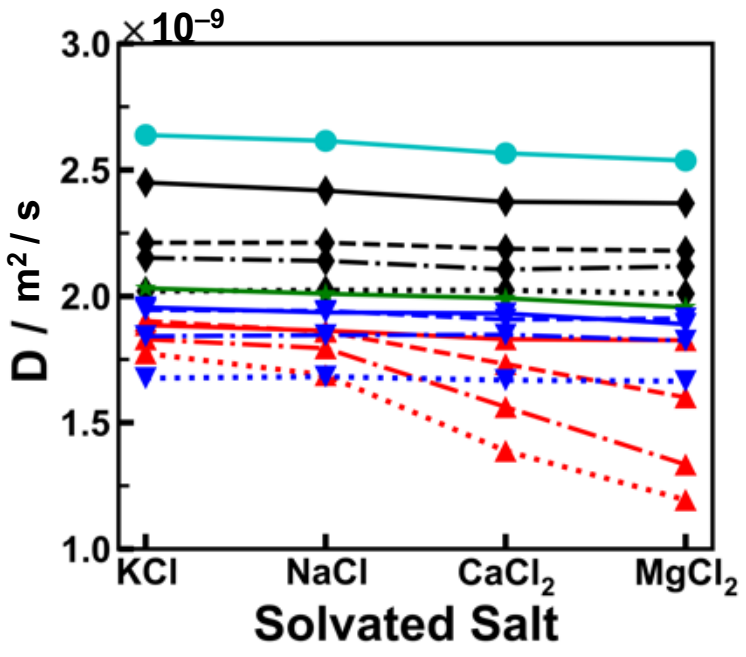

**Figure 3.** Self-diffusion coefficient of water in the whole hydration layer of different DNDs with various surface charges (differentiated by different line styles). Black, blue, red, and green lines represent DND–H, DND–$NH_2$, DND–COOH, and DND–OH particles, respectively. Line styles are the same as those in Figure 2. The cyan line shows the self-diffusion coefficient of water in the bulk region of the neutral DND–H solvated in four different salt solutions.

We have shown the calculated MSDs in Figure 2, in which we observe a linear relationship between MSD and time. Thus, we can readily obtain the self-diffusion coefficients of the DNDs' hydrating water as the slope of MSD plots. In Figure 3, we have compared the effects of different salts on the resulting $D$ values corresponding to various DNDs. We have also shown the self-diffusion coefficients of water in the bulk region of the neutral DND–H solutions far away from the influence of DNDs' surfaces. The numerical values of $D$ as well as the standard error of the linear regression fit are presented in the Supplementary Information (SI) (Table S.1 to Table S.5). We observe couple of interesting trends in Figure 3 that are discussed below.

First, water diffuses the fastest around the uncharged DND–H compared with other uncharged DNDs and it has the closest self-diffusion coefficient to that of the bulk water. It implies that water surrounding DND–H with zero net charge has a bulk-like behavior in terms of the translational motion. We ascribe this behavior to the hydrophobic nature of the DND–H surface, which does not form any HBs with the interfacial water, as opposed to other DNDs in this study. In contrast, surfaces of DND–OH, DND–$NH_2$, and DND–COOH particles are covered with polar groups, which can form HBs with the interfacial water molecules. Thus, they can slow down the motion of the nearby water molecules.

Second, with some few exceptions, the self-diffusion coefficients decrease with increases in the net charge of DNDs. The exceptions are DND–$NH_2$ with +28 net charge solvated in any of salt solutions and DND–COOH with –28 net charge that is solvated in either KCl or NaCl solutions. As opposed to the uncharged DND–$NH_2$, DND–$NH_2$ with +28 net charge has twenty-eight $NH_3^+$ groups on its surfaces. However, it seems they are not sufficient to enhance the formation of strong HBs with the interfacial water.

Third, the negatively charged DND–COOH has a distinct effect on the retardation of water compared with the positively charged DND–H and DND–$NH_2$. In particular, different positive counterions (i.e., $Na^+$, $K^+$, $Ca^{2+}$, $Mg^{2+}$) present in the vicinity of the charged DND–COOHs have contrasting impacts on the mobility of water molecules in the aforementioned hydration shells. Indeed, we can rank them based on their association with the relative slowdown of the water mobility in the whole hydration shell of the charged DND–COOH as

$$Mg^{2+}\ (0.72/4.3) > Ca^{2+}\ (1.0/2.24) > Na^{+}\ (1.02/1.06) > K^{+}\ (1.38/0.59)$$

The numbers inside the parentheses denote ionic radius in Å and the relative charge density of cations, respectively.[6] We also observe the same trend in the self-diffusion coefficients of the bulk water in the corresponding chloride-cation salt solutions.

The aforementioned ranking follows the Hofmeister series[9,31], which suggests that ions with larger charge densities are associated with more pronounced effect on the retardation of water molecules. In particular, if we only compare isovalent ions with each other, water surrounded by the smaller ion species experiences more slowdown in its mobility. We can attribute this effect to the stronger binding energies between smaller cations, which have higher charge densities, and water molecules in the cations' first hydration shell. In Fact, calculations of the entropy of hydration of alkaline metal cations and also halide anions have revealed that $K^+$ and $Cl^-$ ions weakly bind water molecules, while $Na^+$ strongly bind water molecules compared with water-water interactions in the pure water.[69,70] Hence, water molecules are more mobile in the surroundings of $K^+$ cations, compared with those around $Na^+$. Interestingly, it matches quite well with our findings in Figure 3, where we found that the self-diffusion of water around the negatively charged DND–COOH is lower in the presence of adsorbed $Na^+$ cations than that of $K^+$.

To investigate the statistical significance of differences we observed in Figure 3, we have developed the following multiple regression model for self-diffusion coefficient of water molecules in the whole hydration shell of DNDs.

$$D = \beta_0 + \beta_1\, charge + \beta_2\, DND + \beta_3\, DND{:}charge + \beta_4\, salt{:}DND{:}charge \qquad \text{Eq. 8}$$

where $\beta_i$ is the regression coefficient, $DND$ corresponds to one of DND particles (DND–H, DND–COOH, DND–NH2), $charge$ is the number of charges on the DND, and $salt$ is the solvated salt in water (NaCl, KCl, $MgCl_2$, $CaCl_2$). $\beta_3$ and $\beta_4$ are coefficients of the so-called interaction terms between, respectively, ($DND$, $charge$) and ($salt$, $DND$, $charge$). The regression model can help us assess the statistically significance of the effects of DND's surface charges and also the type of the salt on the mobility of DND's interfacial water. Particularly, interaction terms facilitate to investigate whether $charge$ or $salt$ modifies the effect of DND's surface moieties on the interfacial water's self-diffusion coefficient.

**Table 1.** Coefficients of the regression model in Eq. 8. We have assigned DND–COOH and KCl as reference levels to categorical variables $DND$ and $salt$, respectively. Values of $\beta_i$, Standard Error, and confidence interval are expressed in $10^{-9}$ $m^2.sec^{-1}$.

| Variable | $\beta_i$ | Standard Error | p-value | 95% confidence interval |
|---|---|---|---|---|
| **Intercept ($\beta_0$)** | 1.8660 | 0.0082 | **0.0** | (1.8500, 1.8800) |
| $charge$ | -0.0008 | 0.0002 | **0.0** | (-0.0013, -0.0004) |
| $DND$ = **DND–$NH_2$** | 0.1040 | 0.0116 | **0.0** | (0.0811, 0.1270) |
| $DND$ = **DND–H** | 0.5037 | 0.0116 | **0.0** | (0.4810, 0.5270) |
| $DND$ = **DND–$NH_2$ :** $charge$ | 0.0038 | 0.0003 | **0.0** | (0.0031, 0.0044) |
| $DND$ = **DND–H :** $charge$ | 0.0050 | 0.0003 | **0.0** | (0.0044, 0.0056) |
| $salt$ = **NaCl :** $DND$ = **DND–COOH :** $charge$ | -0.0009 | 0.0003 | **0.0** | (-0.0015, -0.0004) |
| $salt$ = **$CaCl_2$ :** $DND$ = **DND–COOH :** $charge$ | -0.0048 | 0.0003 | **0.0** | (-0.0053, -0.0042) |
| $salt$ = **$MgCl_2$ :** $DND$ = **DND–COOH :** $charge$ | -0.0077 | 0.0003 | **0.0** | (-0.0083, -0.0072) |
| $salt$ = **NaCl :** $DND$ = **DND–$NH_2$ :** $charge$ | -0.0001 | 0.0003 | 0.833 | (-0.0006, 0.0005) |
| $salt$ = **$CaCl_2$ :** $DND$ = **DND–$NH_2$ :** $charge$ | 0.0001 | 0.0003 | 0.631 | (-0.0004, 0.0007) |
| $salt$ = **$MgCl_2$ :** $DND$ = **DND–$NH_2$ :** $charge$ | 0.0003 | 0.0003 | 0.309 | (-0.0003, 0.0008) |
| $salt$ = **NaCl :** $DND$ = **DND–$NH_2$ :** $charge$ | 0.0000 | 0.0003 | 0.908 | (-0.0005, 0.0006) |
| $salt$ = **$CaCl_2$ :** $DND$ = **DND–H :** $charge$ | 0.0003 | 0.0003 | 0.319 | (-0.0003, 0.0008) |
| $salt$ = **$MgCl_2$ :** $DND$ = **DND–H :** $charge$ | 0.0003 | 0.0003 | 0.213 | (-0.0002, 0.0009) |

The dataset, which has been used to fit the regression model, contains 240 observations, wherein each combination of $DND$, $charge$, and $salt$ has self-diffusion coefficients calculated from five independent MD simulations. The fitted model has a high adjusted-$R^2$ of 0.972 and its coefficients are listed in Table 1.. Since $DND$ and $salt$ variables are categorical, their corresponding values of $\beta_i$ are given for each specific level of the categorical variable. In this regard, we have assigned, respectively, DND–COOH and KCl as the reference level to the former and latter. Thus, $\beta_0$ corresponds to the self-diffusion coefficient of water in the whole hydration shell of DND–COOH in KCl solution, adjusted for the DND's charges.

With the significance level of 5%, we discard all regression coefficients in Table 1., whose p-values are larger than 0.05. Thus, our model shows that the effect of the solvated salt on the mobility of the DND's interfacial water is statistically significant only for negatively charged DND–COOHs. In contrast, the effect of the amount of DND's surface charges on its interfacial water mobility is not only statistically significant for all types of DNDs, but also this effect gets modified depending on the type of DND. The latter statement is supported by the p-values of the $DND{:}\,charge$ interaction term in Table 1.. These two conclusions match with the trends we previously observed in Figure 3. Based on the explanations above, the final form of the multiple regression model for the self-diffusion coefficient becomes:

$$\begin{aligned} D = {} & 1.866 - 0.0008 \times charge \\ & + 0.1040 \times (\mathrm{DND-NH_2}) + 0.5037 \times (\mathrm{DND-H}) \\ & + 0.0038 \times (\mathrm{DND-NH_2}){:}\, charge + 0.0050 \times (\mathrm{DND-H}){:}\, charge \\ & - 0.0009 \times \mathrm{NaCl}{:}\,(\mathrm{DND-COOH}){:}\, charge \\ & - 0.0048 \times \mathrm{CaCl_2}{:}\,(\mathrm{DND-COOH}){:}\, charge \\ & - 0.0077 \times \mathrm{MaCl_2}{:}\,(\mathrm{DND-COOH}){:}\, charge \end{aligned} \qquad \text{Eq. 9}$$

## 3.2 Rotational dynamics

We now turn our attention to the rotational dynamics of water molecules in the vicinity of DNDs. For this purpose, we have calculated the RCF for dipole and OH vectors of DNDs' interfacial water.

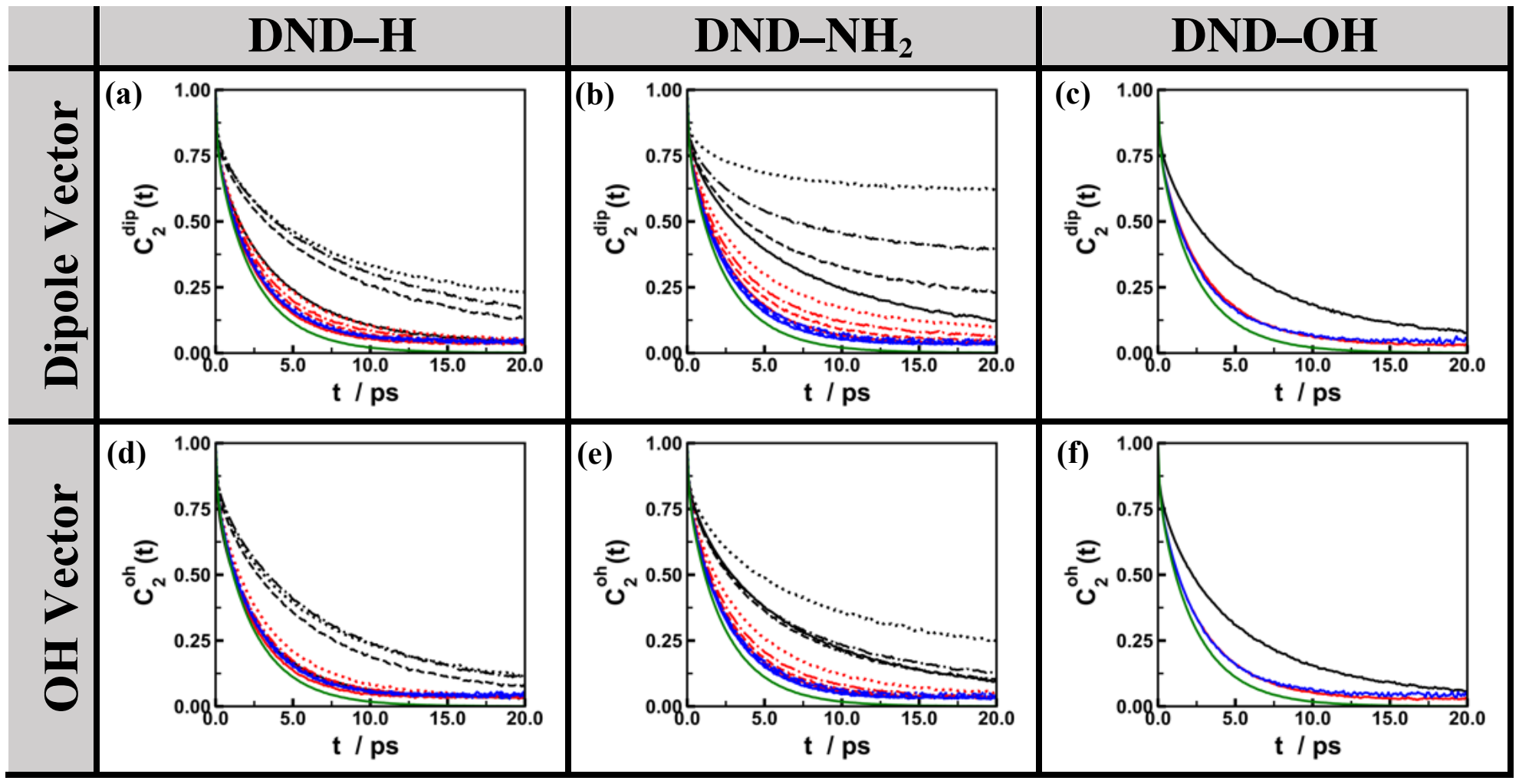

**Figure 4.** RCFs of two water's orientations, dipole and OH vectors, in the first (black line), second (red line), and third (blue line) hydration layers of different DNDs with various surface charges (differentiated by different line styles) that are solvated in NaCl solution. Line styles are the same as those in Figure 2. RCFs of these DNDs in other three salt solutions are almost identical to ones shown here.

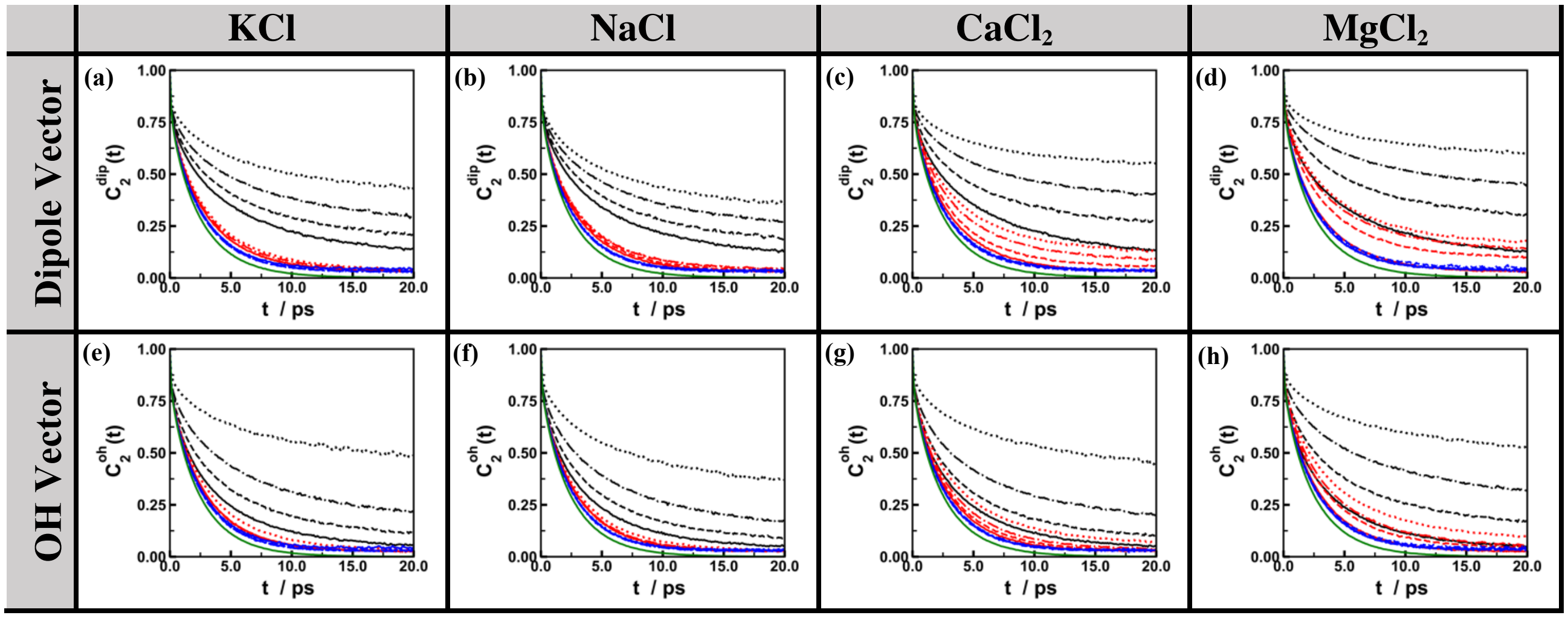

**Figure 5.** Same as Figure 4, but for DND–COOH in four different salt solutions: KCl, NaCl, $CaCl_2$, $MgCl_2$.

We have shown the time evolution of RCFs in Figure 4 for DND–H, DND–$NH_2$, and DND–OH, which are solvated in the NaCl salt solution. Since results for other three salt solutions resemble to those of NaCl salt solution, they have not been included in Figure 4. In contrast, there seems to be an association between the type of the cation in the solution and the decay rate of RCFs for DND–COOHs. Thus, we have demonstrated RCFs for water around DND–COOH in Figure 5, which have been distinguished by the type of the salt in the solution. All plots in these two figures include results for all three hydration layers. We had previously identified these hydration layers elsewhere[60].

We observe three distinct behaviors in the reorientational dynamics of water shown in Figure 4 and Figure 5:

1) Water in the first and second hydration layers of DNDs, except for DND–OH, exhibits anisotropy in the decay rate of its dipole and OH vectors' reorientation. The anisotropic decay appears to be more prominent, as DNDs become more positively or negatively charged, particularly in the case of DND–COOH and DND–$NH_2$ systems.
2) We observe an apparent interaction between the influence of surface charges of DND–COOH and the type of the solvated salt on the RCF decay of both dipole and OH orientations. This trend is similar to what we observed for the translational mobility of DNDs' interfacial water in the previous section. As we pointed out before, the negatively charged DND–COOH attracts cations to its first hydration layer, while positively charged DND–H and DND–$NH_2$ accumulate $Cl^-$ counterions at their interface with water. Since alkali metal and alkaline earth cations have different binding strength with water, they have differing impacts on its various orientational degrees of freedom.
3) All RCFs appear to decay at multiple distinct time scales. They undergo a fast decay, followed by slower relaxations. In particular, when DNDs acquire more surface charges, the slower relaxation visually becomes more obvious.

We use the EWIC model, introduced in Section 2, to quantify the aforementioned different relaxation time scales of RCFs. Furthermore, this model enables us to elucidate the impact of DND's surface chemistries and solution environments on the reorientational dynamics of the interfacial water.

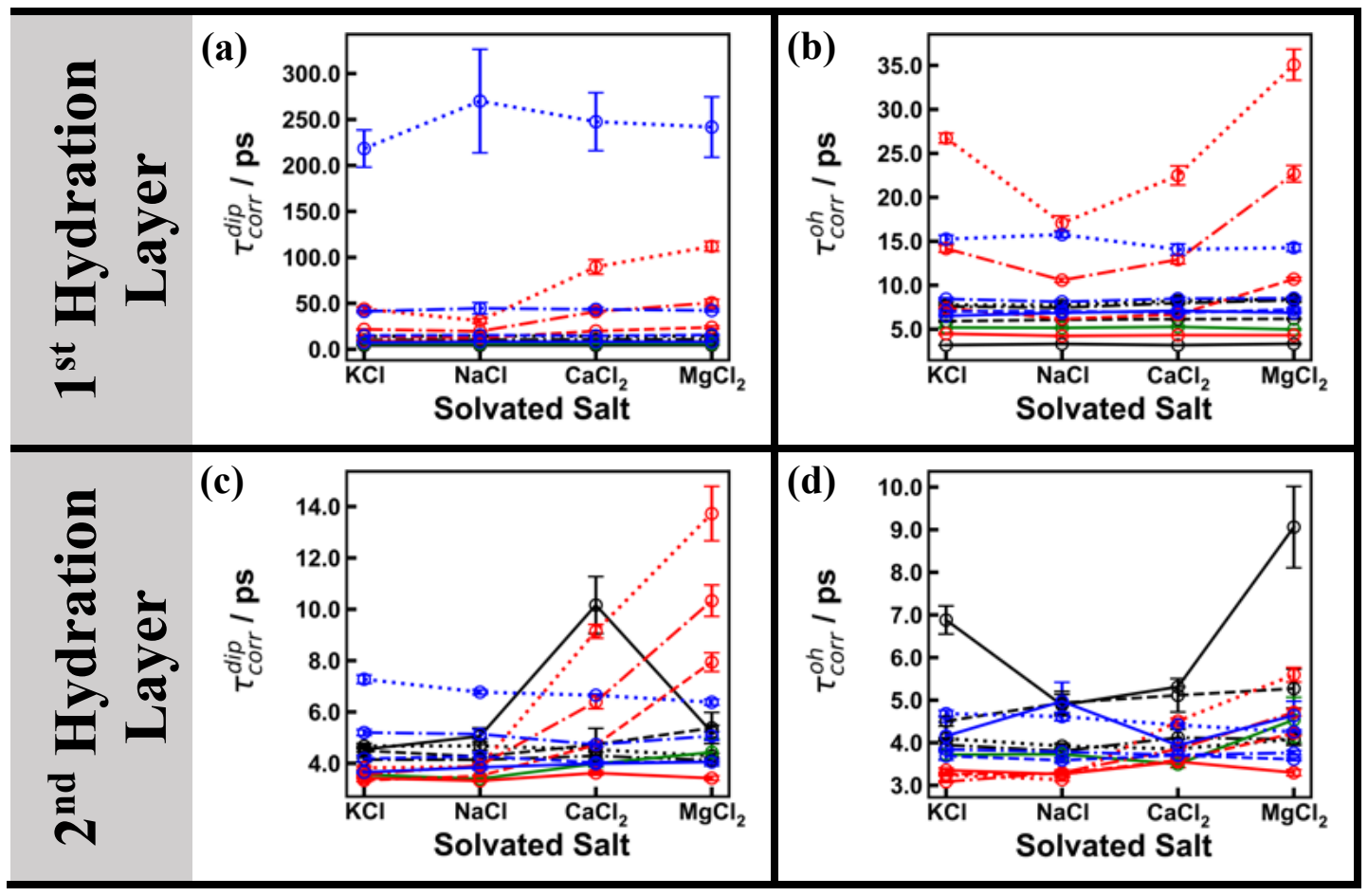

**Figure 6.** Correlation time of RCFs for dipole and OH orientations of water, denoted as $\tau_{corr}^{dip}$ and $\tau_{corr}^{oh}$, in the first and second hydration layers of DNDs. Lines' styles and colors are the same as those in Figure 3.

We have listed relaxation time constants along with other parameters associated with the EWIC model in the SI (Table S.6 through Table S.21). Following statements hold true for both dipole and OH vectors of water. The scale of relaxation time constants, $\tau_{in}$, $\tau_c$, and $\tau_m$ in these tables confirms the existence of three regimes in the orientational relaxation of water that we discussed in Section 2. More specifically, $\tau_{in}$ with sub-picosecond values corresponds to very fast inertial-liberational orientational motion, while $\tau_c$ exhibits values on a larger scale of ~ 1-5 picoseconds. The latter characterizes the constraint rotational diffusion in a cone. In addition, we can observe values as small as ~ 10 picoseconds and as large as ~ 200-400 picoseconds for $\tau_m$, which characterize the final much slower reorientational relaxations. The largest values of $\tau_m$ correspond to reorientational relaxation of water's dipole moment in the first hydration layer of DND–$NH_2$ with +84 absolute charges. It points to the existence of substantial constraints on the water's dipole reorientation in the aforementioned region.

In Figure 6, we have demonstrated the correlation time of RCFs for dipole and OH vectors of water in the first and second hydration layers around DNDs. We have discussed below some interesting patterns that we have found in this figure.

1) The interaction between the absolute net charges of DND–COOHs and the type of solvated salt that we observed in Figure 5 is also reflected in the correlation time of RCF (i.e., $\tau_{corr}$) for dipole (see Figure 6((a), (c)) and OH (see Figure 6((b), (d))) vectors. In fact, we can sort the salts in the following ordering in terms of their association with values of $\tau_{corr}^{dip}$ and $\tau_{corr}^{OH}$:

**NaCl < KCl < $CaCl_2$ < $MgCl_2$**

Smallest $\tau_{corr}^{dip}$ → Largest $\tau_{corr}^{dip}$

**NaCl < $CaCl_2$ < KCl < $MgCl_2$**

Smallest $\tau_{corr}^{oh}$ → Largest $\tau_{corr}^{oh}$

Interestingly, this trend exactly aligns with what we discovered in our previous study[60] about the degree of disorder of water orientations around negatively charged DND–COOHs. More specifically, we observed that water in NaCl and $MgCl_2$ solutions organized themselves, respectively, in the most and the least randomly state around these DNDs.

2) We observe following trends for uncharged DNDs:
    a. $\tau_{corr}^{dip}$ values for water in the first hydration layer of DND–COOH and DND–$NH_2$ are almost identical in all salt solutions except for KCl solution. Moreover, they are 2 and 1.5 times larger than those of DND–H and DND–OH, respectively. In the case of water's OH bond reorientation relaxation, we have found the following ordering among DNDs:

    **DND–H < DND–COOH < DND–OH < DND–$NH_2$**

    Smallest $\tau_{corr}^{oh}$ 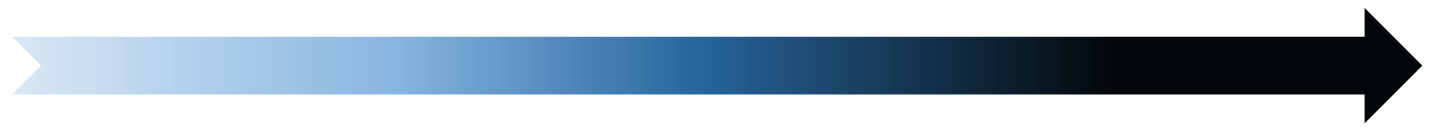 Largest $\tau_{corr}^{oh}$

    We speculate that the driving factor behind this ordering is the nature of HBs between surfaces of these DNDs and water in the first hydration layer. In our previous work[60], we observed that the majority of the aforementioned water molecules point their dipole vector away from facets of the uncharged DND–COOH and thus they act as HB acceptors. In contrast, they orient their dipole towards surfaces of DND–OH and DND–$NH_2$ with zero net charge and become a HB donor. Consequently, OH bonds of water are more engaged with surfaces of the DND–OH and DND–$NH_2$ than those of DND–COOH, which leads to larger $\tau_{corr}^{oh}$ around the former.
    b. For water in the second hydration layer of DNDs with zero net charge, DND–H is associated with the largest values for $\tau_{corr}^{dip}$ and $\tau_{corr}^{oh}$. It could be due to the weakened water-water HBs between the first and the second hydration layers around neutral DNDs with polar surface groups. Indeed, surfaces of these DNDs are decorated with moieties such as COOH, $NH_2$, or OH, which form HBs with water in the first hydration layer. Therefore, it drives the first layer water to break some of its HBs with water in the second hydration layer. In contrast, since the uncharged DND–H is hydrophobic, it drives water in the first hydration layer to form more HBs with water in the second hydration layer.
3) For charged DNDs, observed trends are a little bit more complicated, as discussed below:
    a. $\tau_{corr}^{dip}$ and $\tau_{corr}^{oh}$ corresponding to water in the first hydration layer are smallest for DND–H at all absolute charge values greater than zero. It can be attributed to the less constrained environment for water nearby DND–H, with non-polar surface characteristics, compared to DND–$NH_2$ and DND–COOH.
    b. There seems to be a competition between charged DND–COOH and DND–$NH_2$ particles. In Table 2, we have specified which of these DNDs has won the competition in terms of the largest value for each of $\tau_{corr}^{dip}$ and $\tau_{corr}^{oh}$ corresponding to a specific salt solution and |q| (i.e., the absolute net charges on the DND). The results in Table 2 corresponds to the first hydration layer. This table clearly shows that positively charged DND–$NH_2$ dominates negatively charged DND–COOH in terms of the overall relaxation time for the reorientation of water dipole (i.e., $\tau_{corr}^{dip}$) in the first hydration layer, whereas the reverse is true for $\tau_{corr}^{oh}$. This observation can be attributed to $NH_3^+$ and $COO^-$ species that exist on surfaces of charged DND–$NH_2$ and DND–COOH,

respectively. While $NH_3^+$ acts as a HB donor to water and locks in its dipole, $COO^-$ appears as a HB acceptor from water and hence constrains the reorientation of its OH bonds.

c. There are few exceptions to the overall trend that we just detected in part b above. Indeed, $\tau_{corr}^{dip}$ is larger around DND–COOH than that around DND–$NH_2$ for |q| values of 28 and 56 and in the salt solution of $MgCl_2$. The same is true for the case of |q|=28 and the salt solution of $CaCl_2$. On the other hand, at |q|=28 and in salt solutions of NaCl and $CaCl_2$, it takes longer for water around DND–$NH_2$ to relax its OH bond's initial orientation than that around DND–COOH. We attribute these observations to the significant ordered structure that $Mg^{2+}$ and $Ca^{2+}$ (though relatively with a lesser degree than $Mg^{2+}$) ions impart onto their nearby water by constraining its dipole. Nevertheless, it appears that the strength of water's dipole lockdown imposed by $NH_3^+$ groups of DND–$NH_2$ at |q|=84 is greater than what induced by $Mg^{2+}$ ions surrounding DND–COOH at |q|=84.

d. $\tau_{corr}^{dip}$ for water in the second hydration layer of charged DNDs exhibits the following trend. DND–$NH_2$ is associated with largest values for $\tau_{corr}^{dip}$ at each specific amount of surface charges, if either of NaCl or KCl salts exists in the solution. However, the replacement of either of salts by $CaCl_2$ or $MgCl_2$ salts causes the water in the second hydration layer around DND–COOH to experience the longest relaxation in the reorientation of its dipole, compared with other charged DNDs. This effect can be explained by the substantial constraints that $Mg^{2+}$ or $Ca^{2+}$ cations impose on the dipole vector of their surrounding water. We saw in our previous study[58] that the aforementioned constraints can go beyond the first hydration shell of $Mg^{2+}$ or $Ca^{2+}$.

**Table 2.** Comparison of charged DND–COOH and DND–$NH_2$ particles in different salt solutions and various net absolute charges on the DND (|q|) to determine which one has the dominant effect in slowing down reorientational dynamics of water's dipole and OH orientations in the DND's first hydration layer. The color of a cell specifies the dominant DND.

DND–COOH (red) DND–$NH_2$ (blue)

| | $\tau_{corr}^{dip}$ | | | $\tau_{corr}^{oh}$ | | |
|---|---|---|---|---|---|---|
| **Salt Solution** | \|q\|=28 | \|q\|=56 | \|q\|=84 | \|q\|=28 | \|q\|=56 | \|q\|=84 |
| **KCl** | DND–$NH_2$ | DND–$NH_2$ | DND–$NH_2$ | DND–COOH | DND–COOH | DND–COOH |
| **NaCl** | DND–$NH_2$ | DND–$NH_2$ | DND–$NH_2$ | DND–$NH_2$ | DND–COOH | DND–COOH |
| **$CaCl_2$** | DND–COOH | DND–$NH_2$ | DND–$NH_2$ | DND–$NH_2$ | DND–COOH | DND–COOH |
| **$MgCl_2$** | DND–COOH | DND–COOH | DND–$NH_2$ | DND–COOH | DND–COOH | DND–COOH |

We now aim at digging deeper into mechanisms of the reorientational dynamics of water around DNDs by using the EWIC model as a guide. According to this model, on the one hand, the semi-angle $\theta_{tot}$ of the cone, in which water reorients, and its associated wobbling-in-a-cone (hereafter, wobbling) diffusion coefficient $D_c$ represent the local constraints on water's reorientational dynamics. On the other hand, the diffusion coefficient $D_m$ reflects more global restrictions for complete reorientational relaxation of water. In particular, as Tan *et al.* pointed out[52], $\theta_{tot}$, $D_c$, and $D_m$ for water's OH reorientations reflect two underlying processes in the dynamics of HB networks. The first two parameters reflect local motions within the HB network such as HB stretching or angular vibrations, without breaking HBs. However, the third parameter reveals rearrangements in the HB network in a more global sense. In fact, the more rigid or flexible local HBs are, the narrower or wider the cone of the wobbling water is. Thus, the smaller value of the cone's semi-angle $\theta_{tot}$ lends itself to the slower wobbling diffusion coefficient $D_c$ and vice versa. However, $D_m$ for water's OH is related to the overall structural rearrangements in the HB networks. Therefore, smaller $D_m$ implies that it takes longer to break old HBs and form new ones.

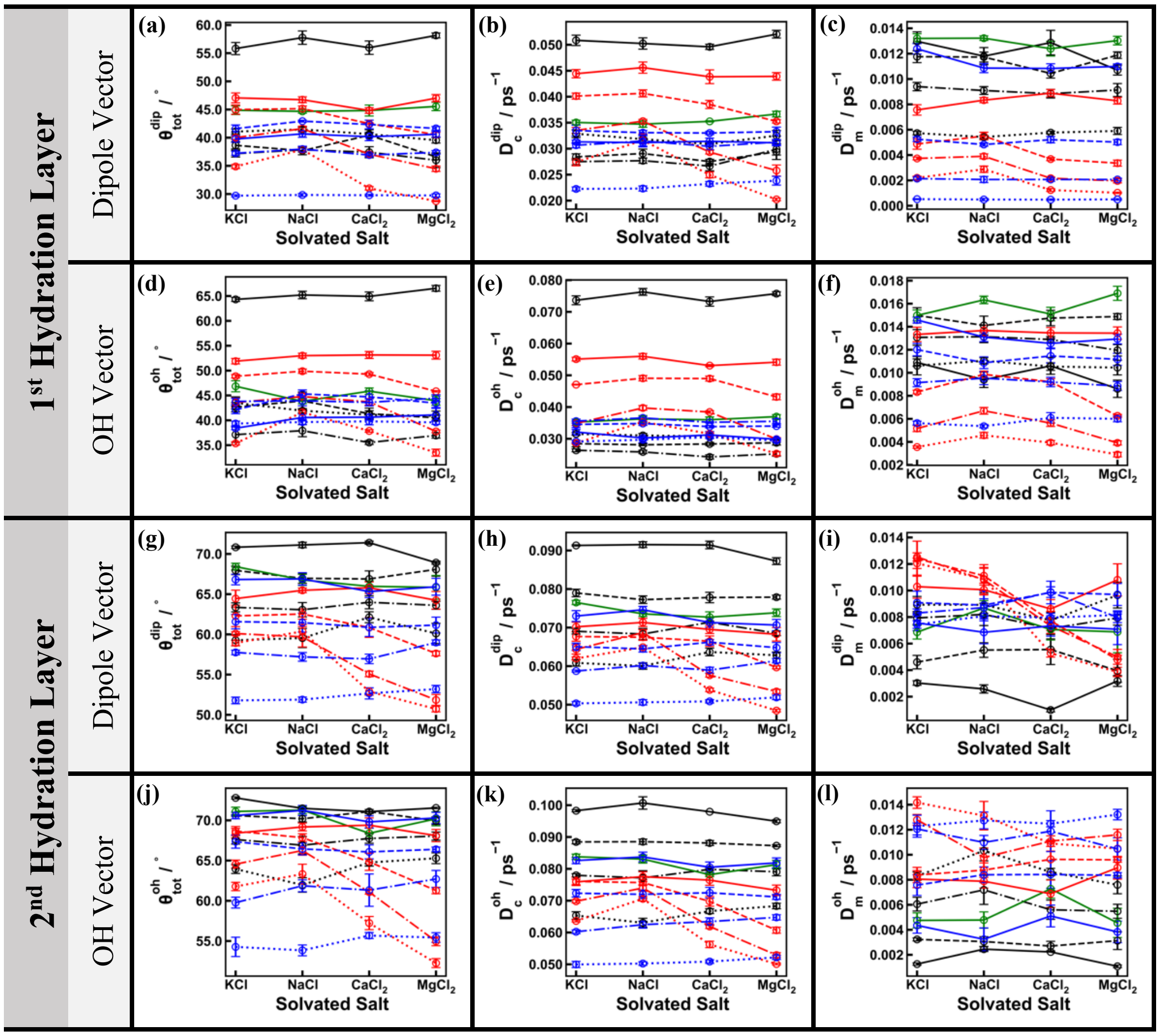

**Figure 7.** Parameters of EWIC model to describe the reorientational dynamics of water's dipole and OH orientations in two hydration layers of DNDs. $\theta_{tot}$, $D_c$, and $D_m$ denote the cone's semi-angle, wobbling diffusion coefficient, and rotational diffusion coefficient, respectively. Lines' styles and colors are the same as those in Figure 3.

With the aforementioned picture of the EWIC model in mind, we embark on analyzing the cone's semi-angle $\theta_{tot}$, wobbling diffusion coefficient $D_c$, and rotational diffusion coefficient $D_m$, all of which are shown in Figure 7. Trends that appear in this figure not only support previous observations for $\tau_{corr}^{dip}$ and $\tau_{corr}^{oh}$, but also reveal new insights that we have summarized below:

1) In both hydration layers, $\theta_{tot}^{dip}$ and $\theta_{tot}^{oh}$ as well as $D_c^{dip}$ and $D_c^{oh}$ corresponding to the uncharged DND–H have the highest values. In other words, the wobbling diffusion in the cone for water's dipole and OH orientations are the fastest nearby DND–H with zero net charge and the cone in which they locally rotate is the widest. It is expected due to the hydrophobic nature of this DND. In addition, since it is uncharged, there is no counterion nearby its surfaces to constrain orientational degrees of freedom of water. However, $D_m$ values corresponding to this DND reveal different reorientational behavior in two hydration layers. We observe that both dipole and OH orientations in the second hydration layer of the uncharged DND–H experience the slowest rotational diffusion out of the cone. It is in utter contrast to corresponding values of $D_m$ in the first hydration layer. It could be due to the enhanced formation of HBs between water molecules in two hydration layers, which we had discussed before.

2) We observe an inverse relationship between the amount of the absolute net charges of DNDs and the values of $\theta_{tot}$ and $D_c$ for both dipole and OH orientations in both hydration layers. Indeed, as the absolute net charges of DNDs increases, the values of these parameters decrease. Furthermore, results for DND–COOH reveal that there appears to be an interaction between the DND's charge and the type of the cation in the salt solution. In particular, we can see that the wobbling diffusion coefficients $D_c^{dip}$ and $D_c^{oh}$ substantially drop for highly charged DND–COOH solvated in $CaCl_2$ and $MgCl_2$ solutions.
3) For charged DNDs, we observe noticeable differences for values of $D_m$ in the first and second hydration layers around different DNDs, which are listed below:
   a. In the first hydration layer around all charged DNDs, both $D_m^{dip}$ and $D_m^{oh}$ decrease with an increase in the DND's net charges.
   b. In contrast, when the DNDs' absolute net charges increase, so do values of $D_m^{dip}$ and $D_m^{oh}$ for water in the second hydration layer of DNDs. However, there are few exceptions to this pattern. First, $D_m^{dip}$ values for water in the second hydration layer of all negatively charged DND–COOH solvated in $CaCl_2$ and $MgCl_2$ salt solutions are smaller compared with the corresponding values of the uncharged DND–COOH. Second, in the same salt solutions just mentioned, water's OH vector undergoes slower out of the cone rotational diffusion in the second hydration layer of DND–COOH with –84 charges than that of DND–COOH with –56 charges. Both of these exceptions can be attributed to the substantially constrained hydration shells around $Ca^{2+}$ and $Mg^{2+}$ cations that accumulate around negatively charged DND–COOH particles.

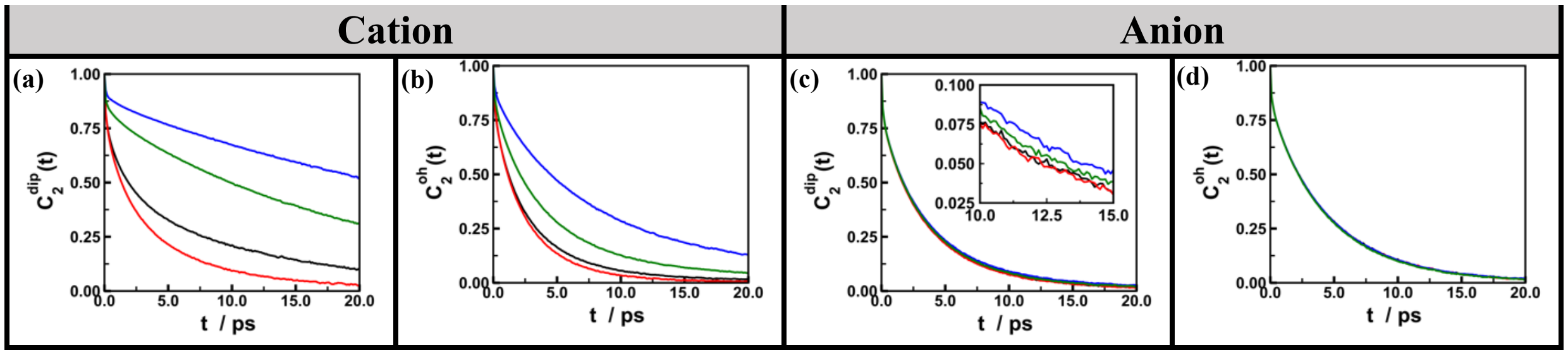

**Figure 8.** RCFs of dipole and OH orientations of water in the first hydration shell of constituent cation and anion of four different salts – KCl (red line), NaCl (black line), $CaCl_2$ (green line), $MgCl_2$ (blue line) – that are solvated in the aqueous solution of the neutral DND–H.

The influence of cations on the reorientational dynamics of water nearby DND–COOH, which we observed above, motivated us to study the same phenomena in the first hydration shell of ions themselves. For this purpose, we calculated RCFs for dipole and OH orientations of water in the first hydration shell of cations and $Cl^-$ anion in the salt solutions of the neutral DND–H. We have demonstrated the results in Figure 8. It is apparent from the decay rate of RCFs in Figure 8 (a-b) that cations slow down the reorientational dynamics of water's dipole moment more intensely than that of its OH vector. The slowdown appears to be more pronounced around the divalent cations than that around the monovalent ones. Conversely, the decay rates of RCFs in Figure 8(c-d) around $Cl^-$ anion seem to be almost identical for both water orientations, although some minor differences exist in RCFs for the dipole vector in four different $Cl^-$ containing salt solutions.

To further elucidate the trends that we observed in Figure 8, we have fitted the tri-exponential function, which is used in the EWIC model, to the corresponding RCFs. The resulting four parameters of this model are shown in Figure 9 for both water orientations and are listed in the SI (Table S.22 and Table S.23). There are a couple of interesting trends in Figure 9 that we note below.

First, Figure 9(a, e) show that the overall relaxation time for dipole vector reorientations around $K^+$, $Na^+$, $Ca^{2+}$, and $Mg^{2+}$ cations are, respectively, 1.5, 2.3, 3.6, and 3.9 times as long as the corresponding values for OH vector. Furthermore, we observe the following ordering in the overall reorientational relaxation time for both water orientations around cations, which follows the Hofmeister series:

$$K^+ < Na^+ << Ca^{2+} < Mg^{2+}$$

Smallest $\tau_{corr}^{dip}$ or $\tau_{corr}^{oh}$

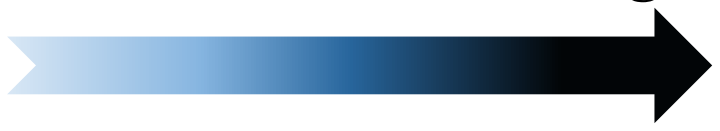

Largest $\tau_{corr}^{dip}$ or $\tau_{corr}^{oh}$

Interestingly, this is exactly the same ordering we found elsewhere[60] for the peak intensity of probability distributions of angles that are formed between each of dipole and OH orientations and the line connecting each of the corresponding cations to the nearby water oxygen atoms. In addition, this trend is in good agreement with findings of Shattuck *et al.*[6] that small, multivalent ions influence the dynamics of the nearby water molecules more profoundly than the large, monovalent ions do. This effect is linked to the higher charge density of the former ions, which in turn they establish stronger electric fields in their surroundings. The strong electric field of $Mg^{2+}$ and $Ca^{2+}$ can be sensed by even more distant water molecules, in contrast to larger, monovalent ions whose influences are mainly restricted to their first hydration shell.

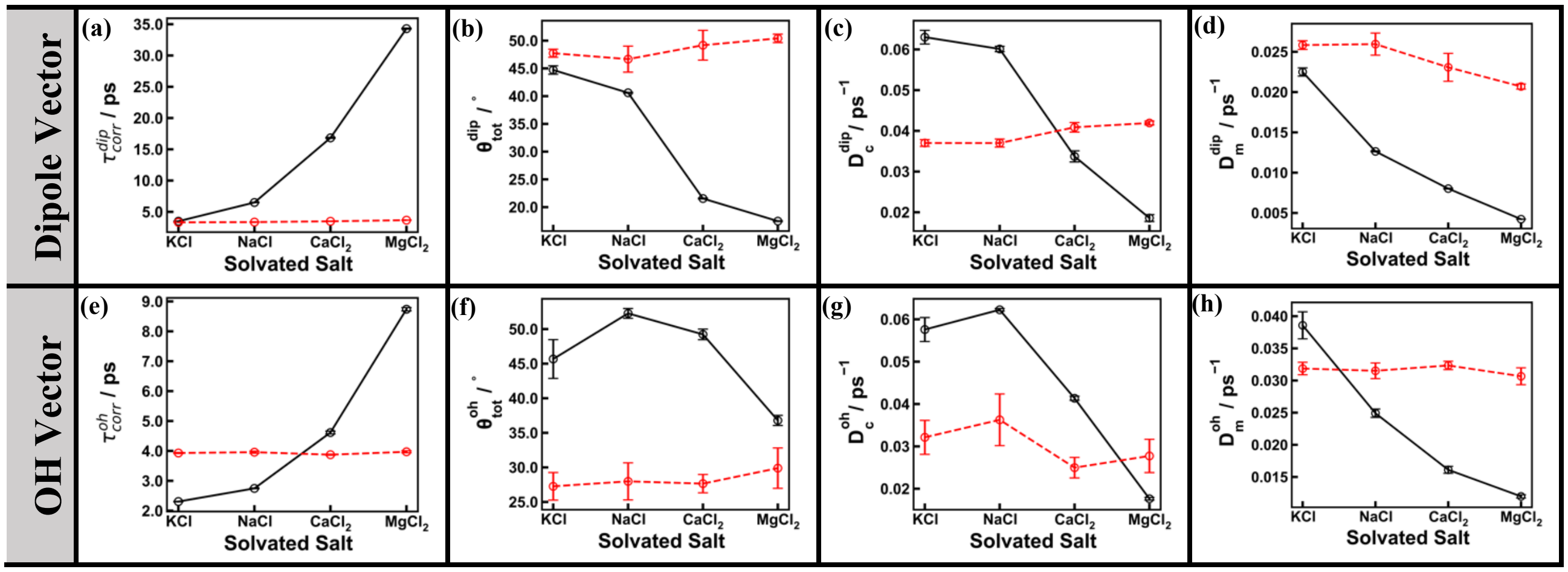

**Figure 9.** Same as Figure 7, but for water in the first hydration shell of constituent cation (black solid line) and anion (red dashed line) of four different salts (KCl, NaCl, $CaCl_2$, $MgCl_2$) that are solvated in the aqueous solution of the neutral DND–H.

Second, the total semi-angle of the cone for the dipole vector of the wobbling water around $Mg^{2+}$ is 17.5°, which is the smallest compared with other cations. Indeed, it is almost 2.6 smaller than the corresponding value for the $K^{+}$ cation. The corresponding values for $Na^{+}$ and $Ca^{2+}$ are 40.6° and 21.5°, respectively. Furthermore, Figure 9(c) reveals that the wobbling diffusion of the dipole vector of water around $Ca^{2+}$ and $Mg^{2+}$ cations is much slower than that around $K^{+}$ and $Na^{+}$ cations. These results show that the wobbling motion of water's dipole in the immediate surrounding of the divalent cations is much more restricted than that of the monovalent ones.

Third, results in Figure 9(f, g) also show more substantially restricted wobbling motion for water's OH vector around $Mg^{2+}$ cation relative to other cations. However, we observe this kind of motion of OH vector around $K^{+}$ is more restricted than that around $Na^{+}$, whereas we saw the reverse behavior for the dipole vector in Figure 9(b, c). The reason behind this observation will be discussed in more details later in this paper.

Four, the rotational diffusion coefficients $D_m^{dip}$ and $D_m^{oh}$ shown in Figure 9(d, h) follow the same ordering, yet in the reverse direction, that we presented for $\tau_{corr}^{dip}$ and $\tau_{corr}^{oh}$ before. These results together with what we observed before for $D_c^{dip}$ and $D_c^{oh}$ lead us to conclude that the reorientation of water's dipole and OH vectors around $Ca^{2+}$ and more notably $Mg^{2+}$ cations are extremely restricted both in their immediate vicinity and further away from them.

Five, values of $\tau_{corr}^{dip}$ around the $Cl^{-}$ anion in KCl, NaCl, $CaCl_2$, and $MgCl_2$ solutions are, respectively, 3.3, 3.4, 3.5, and 3.7 ps. The corresponding values for $\tau_{corr}^{oh}$ are 3.9, 4.0, 3.9, and 4.0 ps, respectively. Thus, as we also saw in Figure 8(c), the overall reorientational relaxations of the dipole vector around the $Cl^{-}$ anion in $CaCl_2$ and $MgCl_2$ solutions are slightly larger than the corresponding values in KCl and NaCl solutions. Figure 9(d) also somehow reflects this trend, where the rotational diffusion of the dipole vector around $Cl^{-}$ anion in $CaCl_2$ and $MgCl_2$ solutions is slower than that in other two solutions. Due to the nature of the rotational diffusion coefficient $D_m$, this effect can be ascribed to the fact that $Mg^{2+}$ and $Ca^{2+}$ (to a lesser degree) can restrict the rotation of water's dipole in larger distances away from themselves relative to two other cations.

Six, averaged over all salt solutions, $\tau_{corr}^{oh}$ of water around $Cl^{-}$ anion is 13% larger than its $\tau_{corr}^{dip}$ around $Cl^{-}$. We attribute this difference to the HB between $Cl^{-}$ and water, which thereby constrains the reorientation of water's OH bond. This statement is also supported by $\theta_{tot}^{dip}$ and $\theta_{tot}^{oh}$ values in Figure 9(b, f), where the latter is almost 50% smaller than the former. That is, constraints induced by $Cl^{-}$– water HB make the wobbling motion of water's OH bond take place in a smaller cone than that of its dipole moment.

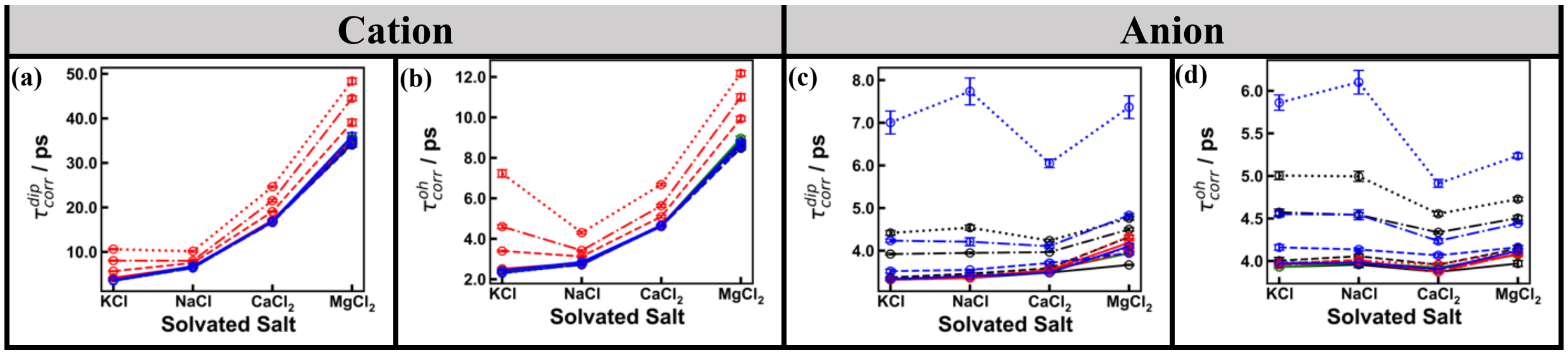

**Figure 10.** Correlation time of RCFs for dipole and OH orientations of water in the first hydration shell of constituent cation and anion of four different salts (KCl, NaCl, $CaCl_2$, $MgCl_2$) that are solvated in the aqueous solution of different DNDs. DNDs are differentiated by different line styles and colors, which are explained in Figure 3.

In Figure 10, we have demonstrated the correlation time of RCFs for water orientations around ions (see SI, Table S.24 to Table S.39 for numerical values). These are different from what we presented in Figure 9 in the sense that Figure 10 also includes results from salt solutions of highly charged DNDs. More specifically, Figure 10 shows enhanced reorientational relaxations of both water orientations around ions in solutions of highly charged DNDs compared with those in uncharged DND solutions. For instance, both $\tau_{corr}^{dip}$ and $\tau_{corr}^{oh}$ values for water around cations in the solution of DND–COOH with –84 charges are noticeably larger than those in the solution of the neutral DND–COOH (see Figure 10(a)). In fact, high concentrations of oppositely charged ions are adsorbed onto the hydration layers of the DND in the former, whereas majority of ions reside further away from the DND's surfaces in the latter.

It is worthwhile to further investigate the aforementioned differences that we observe in the correlation times shown in Figure 10. These differences can be attributed to two factors: (1) the cooperation between surfaces of charged DNDs and interfacial accumulated counterions, (2) relatively high concentrations of counterions in the surrounding of charged DNDs. Prior studies[6,56,71–73] support the likelihood of both factors, though their effects are not necessarily additive. In particular, Tielrooij *et al.* showed that cations and anions act cooperatively with each other and also with the surrounding species (such as polar functional groups on nanoparticles).[32] Depending on the strength of ionic hydrations, the cooperative interactions can manifest different effects on the reorientational dynamics of the nearby water.

We have schematically demonstrated in Figure 11 how cooperative hydration by interfacial counterions and DNDs' charged surface groups manifest themselves in the first hydration layer of DNDs. In the case of positively charged DND–$NH_2$ solutions, $Cl^-$ counterions accumulate in the proximity of DND's $NH_3^+$ functional groups. Thereby, on the one hand, the interfacial water molecules act as hydrogen acceptors and form HBs with $NH_3^+$ moieties as hydrogen donors, which leads to constraining the rotation of water's dipole vector. On the other hand, some of these water molecules donate a hydrogen to $Cl^-$ ions and form HBs with them. Thus, the anions constrain the rotational motion of water's OH bond. In the case of positively charged DND–H solutions, although water does not form HBs with H atoms on DND's surfaces, water's dipole is still constrained by H atoms. However, this constraint is looser than what we described above for DND–$NH_2$ solutions.

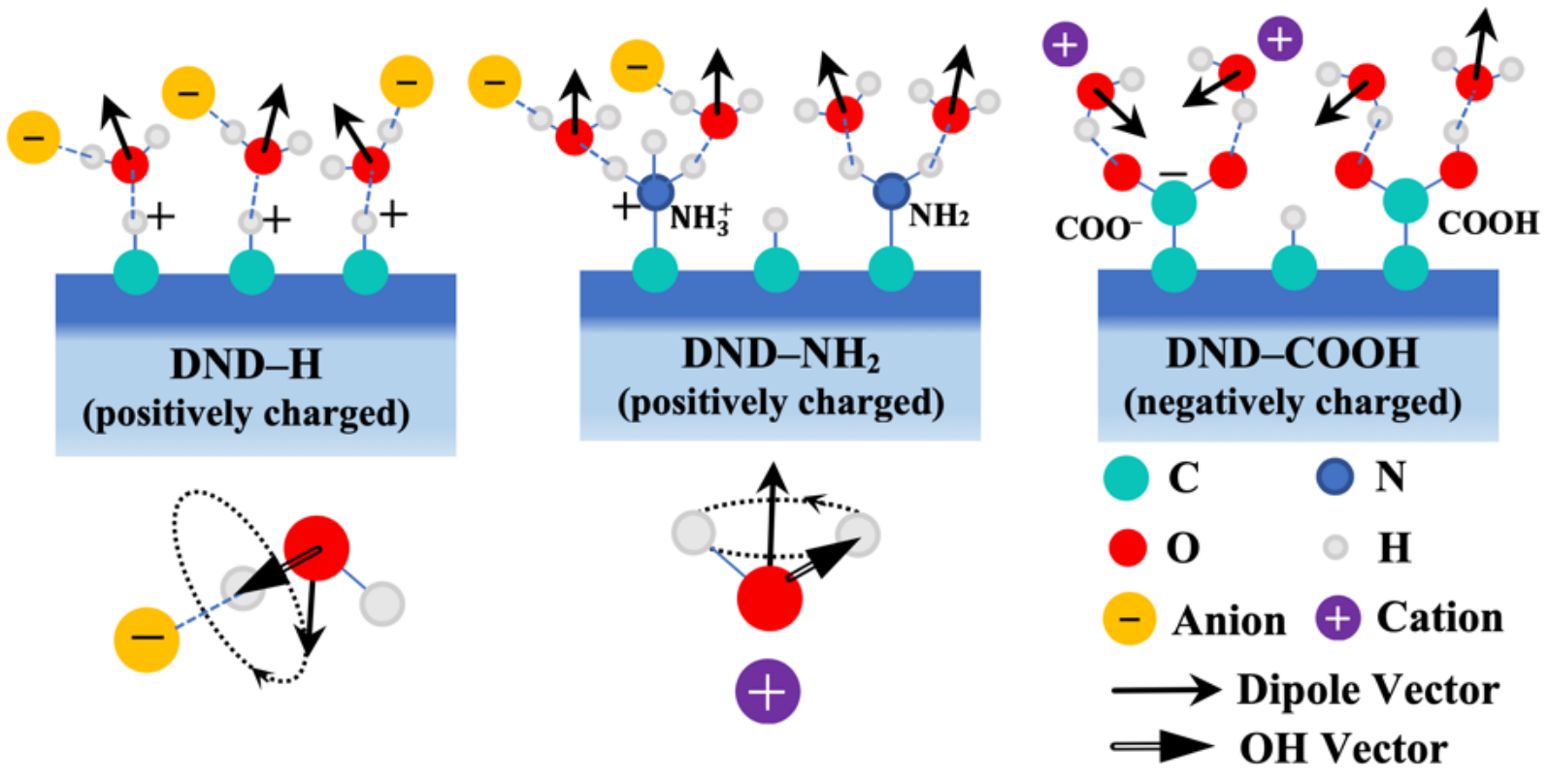

**Figure 11.** A schematic diagram that shows how DND's charged group and the adsorbed counterion on DND' surface can cooperatively influence reorientational dynamics of water. It can manifest in multiple directions, provided that both counterion and DND's charged group have high charge densities. (a)-(c) the first hydration layer of three different charged DNDs, (d) shows that an anion constrains water's OH bond, while its in the first hydration shell of anion, where water's OH bond is constrained by anion.

The cooperative hydration in the first hydration layer of negatively charged DND–COOHs is governed by a different mechanism. The $COO^-$ groups on these DNDs attract cations, which depending on their ionic strengths might be either physically adsorbed on those groups or be separated from those groups by a layer of water. Then, on the one hand, the adsorbed cations lock in the rotation of the interfacial water's dipole vector. On the other hand, some of water molecules in the hydration shell of these cations donate HBs to $COO^-$ groups, which in turn results in constraining the OH bond of those water molecules.

## 4 Conclusion

We have performed MD simulations to study dynamics of the interfacial water in the aqueous salt solutions of DNDs. The goal is to determine how various surface chemistries of DNDs, on the one hand, and ions with different water affinities, on the other hand, influence the translational and rotational motion of water nearby surfaces of DNDs.

We have calculated the self-diffusion coefficient of water, as a signature of its translational mobility, in the whole hydration shell of DNDs from the slope of the MSD vs time plots. We have also developed a multiple regression model to verify if there are any statistically significant associations between the self-diffusion coefficient and various factors in our simulations. These factors include the type of surface functionalization on DNDs, the net charge of DNDs, and the type of solvated salt. Care has been taken to include interactions between these factors in our model. Our model shows significant impacts of all of these factors on the mobility of DNDs' interfacial water, although their degree of influence varies under different conditions. Among all uncharged DNDs, carboxylated DND (DND–COOH) shows the largest slowdown effect on the mobility of the interfacial water, while water nearby the hydrogenated DND (DND–H) experiences the fastest translational motion. The former is explained by the stronger hydrogen bonding between the DND–COOH surfaces and the interfacial water compared with other DNDs, whereas the latter is attributed to the hydrophobicity of DND–H. Furthermore, we have observed that as the absolute net charge of DNDs increases, so does the DNDs' retardation effect on the interfacial water mobility. We ascribe this effect to two interrelated factors: 1) constrained rotational dynamics of the interfacial water, 2) stronger DND–water HBs as a result of the addition of charged polar groups to DND surfaces. The second factor itself also contributes to the first factor. In addition, our regression model shows that there is a statistically significant interaction between the type of the cation in the solution and the amount of DND's charges in solutions of charged carboxylated DNDs. In particular, we have observed $K^+ < Na^+ < Ca^{2+} < Mg^{2+}$ ordering, in terms of the degree of slowdown in the mobility of water in the proximity of the negatively charged DND–COOH, which is solvated in the corresponding chloride solution of these cations. This ordering, which concurs with Hofmeister series, also agrees with experimental observations[74,75]. The significant reduction in the mobility of water associated with divalent cations has been ascribed to their high charge densities[9], which results in the formation of strongly bound hydration shell around them. This is also confirmed by substantially higher residence time of water in the first hydration shell of divalent cations than that of monovalent cations.[60] Consequently, divalent cations carry with themselves their strongly bound water cage[31,76,77], as they move around in the hydration shell of negatively charged DND–COOH. Thereby, it entails the exertion of hydrodynamic frication by the surrounding water molecules. Thus, it eventually leads to significant reduction in the overall mobility of the interfacial water surrounded by $Ca^{2+}$ or $Mg^{2+}$ cations.

We now turn into summarizing our important findings in regard to reorientational dynamics of DNDs' interfacial water. For this purpose, we have presented reorientational correlation functions (RCF) for water's dipole and OH orientations in three hydration layers around DNDs. These hydration layers are defined based on the minima of density plots that we presented in our previous studies[60]. Reorientational dynamics of water in the first two hydration layers are distinct from that of the last one, which is closer to the bulk region. In the former, we have observed that RCFs for both water orientations display three distinct reorientational relaxation modes. The three relaxations modes involve a fast, sub-picosecond decay followed by relaxation on ~ 1-3 picosecond time scale and subsequently a slower relaxation on 10s-100s of picoseconds time scale. However, the RCF in the third hydration layer only displayed the aforementioned first and second relaxation modes, which is similar to what has been reported for bulk water[52].

We have employed the extended wobbling-in-a-cone (EWIC) model to characterize RCF decays in the two closest hydration layers to DNDs' surfaces and thereby to elucidate the mechanism governing the rotational dynamics of the interfacial water. In this model, the first and second RCF relaxations correspond to liberational and wobbling diffusion in hypothetical cones and the last slower relaxation is related to the rotational diffusion of the whole frame of water, which leads to the complete orientational randomization. Our results show that the positively charged DND–$NH_2$ and the negatively charged DND–COOH more substantially influence the orientational dynamics of water in their first hydration layers, compared with other DNDs. In general, in the first hydration layer, the former more predominantly retards the rotational dynamics of water's dipole, while the latter has dominant effects on slowing down the dynamics of water's OH rotation. We attribute these effects to HBs that $NH_3^+$ on DND–$NH_2$ and $COO^-$ on DND–COOH form with the interfacial water. The former is a hydrogen donor and hence lock in the rotation of water's dipole, whereas the latter is a hydrogen acceptor and thus constrains the reorientation of water's OH bond.

Similar to the translational dynamics, we have also observed specific cation effects on the rotational dynamics of both water's orientations in hydration layers of negatively charged DND–COOH. In this regard, we have learned that cations can be ranked as $Na^+ < K^+ < Ca^{2+} < Mg^{2+}$ and $Na^+ < Ca^{2+} < K^+ < Mg^{2+}$. These rankings reflect the degree of slowdown in the orientational dynamics of, respectively, dipole and OH vectors of water in the first hydration layer of charged DND–COOH. Since water is relatively weakly bound to $K^+$ compared with other cations, it can form HBs with neighboring moieties (either water or functional groups on DND–COOH). Thus, it explains why water's OH reorientation is more constrained in the presence of $K^+$ than that of $Na^+$ and $Ca^{2+}$ cations. It also agrees well with experimental observations[6] that small charge density cations are structure-makers outside of their first hydration shell. However, $Mg^{2+}$ binds water so tightly to its hydration shell that both water orientations are substantially constrained.

Our observations have led us to put forward a model for the reorientational dynamics of water hydrating charged DNDs that are surrounded by oppositely charged ions. In this model, which has been depicted in Figure 11, water hydrates charged DNDs in cooperation with adsorbed counterions. Thereby, the charged and polar functional groups on DNDs lock in either of water's dipole or OH orientations and counterions constrain the other orientation which has been pushed away from DNDs' surfaces. We suggest that the effect of the charged DND–counterion cooperative hydration on the dynamics of the interfacial water depends on three factors. These factors, which have also been reported in other studies[9,32], are the charge density of counterions, the concentration of counterions, and the strength of DND–water HB. We have observed the manifestations of the first two factors in both translational and reorientational dynamics of the interfacial water around negatively charged DND–COOH, surrounded by positively charged counterions. In particular, $Mg^{2+}$ with its strongly bound and highly immobilized hydration shells, imparted the slowest dynamics in the interfacial water, compared with other cations. Interestingly, higher concentration of $Mg^{2+}$ reinforced this effect. The role of the third factor can be explained by energy cost associated with breaking the HB between the interfacial water and the DND's surface polar groups. We particularly observed the effect of this factor in the substantially retarded dipolar reorientational dynamics of water close to the DND covered with 84 $NH_3^+$ groups.

In summary, we have found that DNDs in cooperation with solvated ions have substantial impacts on both translational and rotational dynamics of DNDs' interfacial water. These results have implications for the dynamics of hydrogen bonding network of water around DNDs, which have significant impacts on performance of DNDs as fluorescent agents in bioimaging applications[46]. In addition, our study paves the way to further investigate how the slowdown of the interfacial water's translational and rotational dynamics play a role in the colloidal stability of aqueous solution of DNDs.

## Supplementary Information (SI)

This section contains tables of parameters introduced in the main body of the paper for the translational and the reorientational motion of water around various detonation nanodiamonds (DNDs) and ions in aqueous solutions. In Section S.1, the 95% confidence interval (CI) for the average diffusion coefficient of water in the whole hydration shell of each distinct DND in a specific salt solution is presented. We have calculated the CI using the bootstrap percentile method[1], where the bootstrap distribution was obtained by resampling the initial sample data 10,000 times with replacement. We have computed the average value of all parameters in this appendix from five independent MD trajectories. Thus, the sample data for each distinct DND in a particular salt solution have five data points.

### S.1. Translational dynamics

As it was mentioned in Section 2, the self-diffusion coefficient $D$ is obtained from the slope of the Mean Square Displacement (MSD) vs. time plot. We estimated this slope from a linear regression model fitted to the plot. The Standard Error (SE) is provided in Table S.1–Table S.5. for the estimate of $D$ corresponding to each of five independent MD trajectories.

**Table S.1.** The self-diffusion coefficient $D$ of water in the whole hydration shell of DND–H with various surface chemistries. $D_i$ ($i$=1, 2, …, 5) corresponds to the $i$th MD trajectory and $D_{mean}$ is the average value. Values of $D$, SE, and CI are expressed in $10^{-9}$ $m^2.sec^{-1}$.

| DND's Charge | Dissolved Salt | $D_1$ (SE) | $D_2$ (SE) | $D_3$ (SE) | $D_4$ (SE) | $D_5$ (SE) | $D_{mean}$ (CI) |
|---|---|---|---|---|---|---|---|
| q = 0 | KCl | 2.45 (4.98E-3) | 2.44 (3.71E-3) | 2.46 (4.33E-3) | 2.46 (4.52E-3) | 2.44 (3.87E-3) | 2.45 (2.442, 2.458) |
| | NaCl | 2.45 (4.29E-3) | 2.38 (3.71E-3) | 2.41 (4.08E-3) | 2.41 (3.61E-3) | 2.44 (3.75E-3) | 2.418 (2.398, 2.438) |
| | $CaCl_2$ | 2.32 (2.78E-3) | 2.39 (3.37E-3) | 2.39 (3.38E-3) | 2.4 (3.75E-3) | 2.37 (3.86E-3) | 2.374 (2.346, 2.394) |
| | $MgCl_2$ | 2.37 (2.88E-3) | 2.37 (3.44E-3) | 2.34 (4.07E-3) | 2.37 (3.34E-3) | 2.39 (3.50E-3) | 2.368 (2.352, 2.382) |
| q = +28 | KCl | 2.23 (4.75E-3) | 2.22 (4.76E-3) | 2.2 (4.30E-3) | 2.21 (4.83E-3) | 2.2 (5.44E-3) | 2.212 (2.202, 2.222) |
| | NaCl | 2.21 (4.94E-3) | 2.25 (5.21E-3) | 2.21 (5.70E-3) | 2.19 (4.76E-3) | 2.2 (4.98E-3) | 2.212 (2.198, 2.232) |
| | $CaCl_2$ | 2.18 (4.59E-3) | 2.16 (4.49E-3) | 2.19 (4.93E-3) | 2.17 (4.93E-3) | 2.24 (5.11E-3) | 2.188 (2.168, 2.214) |
| | $MgCl_2$ | 2.18 (4.24E-3) | 2.14 (4.48E-3) | 2.15 (4.27E-3) | 2.23 (5.21E-3) | 2.2 (4.40E-3) | 2.18 (2.152, 2.208) |
| q = +56 | KCl | 2.14 (4.97E-3) | 2.15 (6.08E-3) | 2.15 (6.13E-3) | 2.15 (5.50E-3) | 2.17 (6.64E-3) | 2.152 (2.144, 2.162) |
| | NaCl | 2.13 (5.60E-3) | 2.19 (5.66E-3) | 2.13 (5.07E-3) | 2.12 (5.19E-3) | 2.13 (5.23E-3) | 2.14 (2.124, 2.166) |
| | $CaCl_2$ | 2.11 (4.61E-3) | 2.1 (4.54E-3) | 2.11 (4.63E-3) | 2.1 (4.76E-3) | 2.11 (4.49E-3) | 2.106 (2.102, 2.110) |
| | **$MgCl_2$** | 2.09 (4.60E-3) | 2.14 (5.33E-3) | 2.13 (5.02E-3) | 2.13 (5.24E-3) | 2.1 (5.62E-3) | 2.118 (2.100, 2.134) |
| **q = +84** | **KCl** | 2.03 (6.13E-3) | 2.05 (5.43E-3) | 2 (5.79E-3) | 2.03 (5.48E-3) | 1.99 (5.39E-3) | 2.02 (2.000, 2.038) |
| | **NaCl** | 2.04 (5.41E-3) | 2.02 (5.82E-3) | 2 (5.00E-3) | 2.02 (5.74E-3) | 2.04 (5.88E-3) | 2.024 (2.012, 2.036) |
| | **$CaCl_2$** | 2.01 (5.44E-3) | 2.03 (5.41E-3) | 2.06 (5.52E-3) | 2.02 (6.19E-3) | 2 (5.48E-3) | 2.024 (2.008, 2.044) |
| | **$MgCl_2$** | 2 (4.99E-3) | 1.98 (4.87E-3) | 2 (5.39E-3) | 2.05 (5.13E-3) | 2.02 (5.03E-3) | 2.01 (1.992, 2.032) |

**Table S.2.** Same as Table S.1, but for the case of DND–$NH_2$.

| **DND's Charge** | **Dissolved Salt** | ***$D_1$* (SE)** | ***$D_2$* (SE)** | ***$D_3$* (SE)** | ***$D_4$* (SE)** | ***$D_5$* (SE)** | ***$D_{mean}$* (CI)** |
|---|---|---|---|---|---|---|---|
| **q = 0** | **KCl** | 1.95 (5.85E-03) | 1.93 (6.26E-03) | 1.94 (5.47E-03) | 1.98 (6.16E-03) | 1.98 (6.36E-03) | 1.956 (1.938, 1.974) |
| | **NaCl** | 1.94 (7.02E-03) | 1.93 (5.90E-03) | 1.97 (6.56E-03) | 1.91 (5.51E-03) | 1.93 (5.51E-03) | 1.936 (1.920, 1.954) |
| | **$CaCl_2$** | 1.91 (5.94E-03) | 1.92 (6.04E-03) | 1.95 (6.23E-03) | 1.95 (6.41E-03) | 1.93 (6.00E-03) | 1.932 (1.918, 1.946) |
| | **$MgCl_2$** | 1.87 (6.09E-03) | 1.92 (5.91E-03) | 1.87 (5.37E-03) | 1.9 (5.38E-03) | 1.89 (6.15E-03) | 1.89 (1.874, 1.906) |
| **q = +28** | **KCl** | 1.99 (7.38E-03) | 1.93 (6.12E-03) | 1.94 (5.96E-03) | 1.94 (6.89E-03) | 1.93 (5.78E-03) | 1.946 (1.932, 1.968) |
| | **NaCl** | 1.96 (6.15E-03) | 1.96 (5.94E-03) | 1.95 (5.82E-03) | 1.93 (6.13E-03) | 1.91 (6.11E-03) | 1.942 (1.924, 1.958) |
| | **$CaCl_2$** | 1.92 (5.56E-03) | 1.93 (6.41E-03) | 1.91 (5.88E-03) | 1.9 (6.34E-03) | 1.88 (4.89E-03) | 1.908 (1.892, 1.922) |
| | **$MgCl_2$** | 1.91 (6.36E-03) | 1.89 (5.74E-03) | 1.92 (6.24E-03) | 1.94 (5.41E-03) | 1.9 (5.33E-03) | 1.912 (1.898, 1.928) |
| **q = +56** | **KCl** | 1.85 (6.51E-03) | 1.85 (5.86E-03) | 1.81 (5.29E-03) | 1.87 (6.80E-03) | 1.83 (4.90E-03) | 1.842 (1.826, 1.858) |
| | **NaCl** | 1.87 (6.15E-03) | 1.85 (6.39E-03) | 1.85 (5.78E-03) | 1.84 (5.52E-03) | 1.82 (5.75E-03) | 1.846 (1.832, 1.860) |
| | **$CaCl_2$** | 1.88 (6.36E-03) | 1.86 (5.21E-03) | 1.81 (5.09E-03) | 1.84 (5.76E-03) | 1.85 (6.03E-03) | 1.848 (1.828, 1.868) |
| | **$MgCl_2$** | 1.83 (5.49E-03) | 1.81 (5.31E-03) | 1.83 (6.33E-03) | 1.81 (5.31E-03) | 1.84 (5.65E-03) | 1.824 (1.814, 1.834) |
| **q = +84** | **KCl** | 1.68 (6.62E-03) | 1.71 (6.30E-03) | 1.66 (6.40E-03) | 1.65 (5.67E-03) | 1.68 (6.29E-03) | 1.676 (1.658, 1.694) |
| | **NaCl** | 1.67 (6.03E-03) | 1.65 (5.11E-03) | 1.7 (4.85E-03) | 1.71 (6.10E-03) | 1.68 (6.87E-03) | 1.682 (1.664, 1.700) |
| | **$CaCl_2$** | 1.66 | 1.66 | 1.69 | 1.67 | 1.66 | 1.668 |
| | | (5.76E-03) | (5.62E-03) | (6.38E-03) | (5.19E-03) | (5.53E-03) | (1.660, 1.680) |
| | **$MgCl_2$** | 1.67<br>(5.42E-03) | 1.68<br>(5.22E-03) | 1.65<br>(5.65E-03) | 1.66<br>(5.73E-03) | 1.66<br>(5.42E-03) | 1.664<br>(1.656, 1.674) |

**Table S.3.** Same as Table S.1, but for the case of DND–COOH.

| **DND's Charge** | **Dissolved Salt** | ***$D_1$*** **(SE)** | ***$D_2$*** **(SE)** | ***$D_3$*** **(SE)** | ***$D_4$*** **(SE)** | ***$D_5$*** **(SE)** | ***$D_{mean}$*** **(CI)** |
|---|---|---|---|---|---|---|---|
| **q = 0** | **KCl** | 1.9<br>(7.34E-03) | 1.9<br>(8.11E-03) | 1.87<br>(7.26E-03) | 1.91<br>(6.81E-03) | 1.85<br>(6.29E-03) | 1.886<br>(1.866, 1.904) |
| | **NaCl** | 1.86<br>(6.97E-03) | 1.86<br>(6.45E-03) | 1.84<br>(5.99E-03) | 1.88<br>(6.49E-03) | 1.88<br>(6.60E-03) | 1.864<br>(1.852, 1.876) |
| | **$CaCl_2$** | 1.84<br>(7.30E-03) | 1.81<br>(6.65E-03) | 1.82<br>(7.27E-03) | 1.82<br>(6.06E-03) | 1.86<br>(7.57E-03) | 1.83<br>(1.816, 1.848) |
| | **$MgCl_2$** | 1.84<br>(6.36E-03) | 1.8<br>(6.99E-03) | 1.83<br>(6.93E-03) | 1.82<br>(6.28E-03) | 1.84<br>(6.43E-03) | 1.826<br>(1.812, 1.838) |
| **q = –28** | **KCl** | 1.94<br>(6.73E-03) | 1.92<br>(6.85E-03) | 1.91<br>(6.64E-03) | 1.89<br>(6.12E-03) | 1.85<br>(6.48E-03) | 1.902<br>(1.874, 1.926) |
| | **NaCl** | 1.82<br>(5.99E-03) | 1.86<br>(6.90E-03) | 1.85<br>(6.53E-03) | 1.9<br>(6.77E-03) | 1.87<br>(6.65E-03) | 1.86<br>(1.836, 1.882) |
| | **$CaCl_2$** | 1.72<br>(6.06E-03) | 1.73<br>(6.46E-03) | 1.73<br>(6.06E-03) | 1.74<br>(5.97E-03) | 1.74<br>(6.31E-03) | 1.732<br>(1.726, 1.738) |
| | **$MgCl_2$** | 1.6<br>(5.52E-03) | 1.6<br>(5.76E-03) | 1.62<br>(6.04E-03) | 1.58<br>(5.61E-03) | 1.6<br>(6.17E-03) | 1.6<br>(1.588, 1.612) |
| **q = –56** | **KCl** | 1.83<br>(6.44E-03) | 1.85<br>(7.76E-03) | 1.83<br>(6.75E-03) | 1.8<br>(6.06E-03) | 1.84<br>(8.00E-03) | 1.83<br>(1.814, 1.842) |
| | **NaCl** | 1.8<br>(6.61E-03) | 1.78<br>(7.22E-03) | 1.81<br>(6.48E-03) | 1.81<br>(6.42E-03) | 1.77<br>(6.72E-03) | 1.794<br>(1.780, 1.808) |
| | **$CaCl_2$** | 1.61<br>(6.57E-03) | 1.53<br>(6.55E-03) | 1.57<br>(5.70E-03) | 1.54<br>(6.04E-03) | 1.56<br>(6.17E-03) | 1.562<br>(1.540, 1.588) |
| | **$MgCl_2$** | 1.34<br>(5.74E-03) | 1.32<br>(4.95E-03) | 1.31<br>(5.05E-03) | 1.34<br>(4.83E-03) | 1.36<br>(5.55E-03) | 1.334<br>(1.318, 1.348) |
| **q = –84** | **KCl** | 1.76<br>(6.23E-03) | 1.8<br>(7.21E-03) | 1.76<br>(6.56E-03) | 1.79<br>(6.88E-03) | 1.76<br>(6.81E-03) | 1.774<br>(1.760, 1.790) |
| | **NaCl** | 1.69<br>(6.60E-03) | 1.68<br>(5.35E-03) | 1.7<br>(6.95E-03) | 1.67<br>(6.27E-03) | 1.7<br>(6.73E-03) | 1.688<br>(1.678, 1.698) |
| | **$CaCl_2$** | 1.39<br>(5.69E-03) | 1.39<br>(5.57E-03) | 1.37<br>(5.32E-03) | 1.38<br>(6.33E-03) | 1.41<br>(5.78E-03) | 1.388<br>(1.376, 1.400) |
| | **$MgCl_2$** | 1.2<br>(5.01E-03) | 1.19<br>(4.80E-03) | 1.18<br>(4.68E-03) | 1.21<br>(5.37E-03) | 1.19<br>(5.03E-03) | 1.194<br>(1.186, 1.204) |

**Table S.4.** Same as Table S.1, but for the case of DND–OH.

| **DND's Charge** | **Dissolved Salt** | ***$D_1$*** **(SE)** | ***$D_2$*** **(SE)** | ***$D_3$*** **(SE)** | ***$D_4$*** **(SE)** | ***$D_5$*** **(SE)** | ***$D_{mean}$*** **(CI)** |
|---|---|---|---|---|---|---|---|
| **q = 0** | **KCl** | 2.01 | 2.07 | 2.05 | 2.01 | 2.02 | 2.032 |
| | | (5.89E-03) | (7.01E-03) | (5.41E-03) | (4.80E-03) | (5.92E-03) | (2.012, 2.054) |
| | **NaCl** | 2.04<br>(6.37E-03) | 2<br>(5.72E-03) | 2<br>(5.81E-03) | 2<br>(5.18E-03) | 2.01<br>(5.89E-03) | 2.01<br>(2.000, 2.026) |
| | **$CaCl_2$** | 2.02<br>(5.29E-03) | 2<br>(6.45E-03) | 1.98<br>(6.59E-03) | 1.98<br>(5.73E-03) | 1.98<br>(5.40E-03) | 1.992<br>(1.980, 2.008) |
| | **$MgCl_2$** | 1.93<br>(4.99E-03) | 1.99<br>(5.89E-03) | 1.98<br>(6.30E-03) | 1.92<br>(5.34E-03) | 1.96<br>(6.05E-03) | 1.956<br>(1.932, 1.980) |

**Table S.5.** Same as Table S.1, but for the case of the bulk water, which corresponds to the region outside the whole hydration shell of the neutral DND–H in different salt solutions.

| Dissolved Salt | $D_1$ (SE) | $D_2$ (SE) | $D_3$ (SE) | $D_4$ (SE) | $D_5$ (SE) | $D_{mean}$ (CI) |
|---|---|---|---|---|---|---|
| **KCl** | 2.68<br>(1.26E-03) | 2.67<br>(1.56E-03) | 2.65<br>(1.56E-03) | 2.66<br>(1.47E-03) | 2.65<br>(1.46E-03) | 2.662<br>(2.652, 2.672) |
| **NaCl** | 2.65<br>(1.22E-03) | 2.63<br>(1.50E-03) | 2.62<br>(1.66E-03) | 2.63<br>(1.32E-03) | 2.65<br>(1.52E-03) | 2.636<br>(2.626, 2.646) |
| **$CaCl_2$** | 2.61<br>(1.61E-03) | 2.56<br>(1.55E-03) | 2.6<br>(1.38E-03) | 2.59<br>(1.44E-03) | 2.58<br>(1.60E-03) | 2.588<br>(2.572, 2.602) |
| **$MgCl_2$** | 2.53<br>(1.59E-03) | 2.57<br>(1.38E-03) | 2.58<br>(9.97E-04) | 2.56<br>(1.20E-03) | 2.57<br>(1.64E-03) | 2.562<br>(2.546, 2.574) |

## S.2. Reorientational dynamics

In this section parameters of the Extended Wobble-In-Cone (EWIC) model for the reorientational dynamics of the dipole moment and OH bonds of water in the two closest hydration layers to the facets of DNDs as well as the first hydration shell of ions are presented. We defined the EWIC model in Section 2.2 to study the reorientational dynamics of water. The values throughout this section are averaged over the five independent MD trajectories. Due to the space limit, the bootstrapped CI for these point estimates is not presented here.

### S.2.1. **First hydration layer of DNDs**

### S.2.1.1. Water's dipole reorientation parameters

**Table S.6.** The EWIC parameters for the dipole reorientations of water in the first hydration layer of DND–H with various surface chemistries.

| DND's Charge | Dissolved Salt | $\tau_{in}^{dip}$ / ps | $\tau_{c}^{dip}$ / ps | $\tau_{m}^{dip}$ / ps | $\tau_{corr}^{dip}$ / ps | $\theta_{tot}^{dip}$ / ° | $d_{c}^{dip}$ / ps$^{-1}$ | $d_{m}^{dip}$ / ps$^{-1}$ |
|---|---|---|---|---|---|---|---|---|
| **q = 0** | **KCl** | 0.10364 | 3.86686 | 13.10119 | 4.16814 | 55.81 | 0.05085 | 0.01297 |
| | **NaCl** | 0.10806 | 4.07464 | 14.35162 | 4.22332 | 57.75 | 0.05024 | 0.0118 |
| | **$CaCl_2$** | 0.10728 | 3.97407 | 13.2983 | 4.20704 | 55.97 | 0.0496 | 0.01287 |
| | **$MgCl_2$** | 0.10925 | 3.97346 | 15.67982 | 4.40647 | 58.14 | 0.05204 | 0.0107 |
| **q = +28** | **KCl** | 0.09897 | 4.0275 | 14.2902 | 7.86731 | 38.64 | 0.02842 | 0.01176 |
| | **NaCl** | 0.0783 | 3.79346 | 14.28877 | 8.0469 | 37.69 | 0.02906 | 0.01173 |
| | **$CaCl_2$** | 0.0997 | 4.4814 | 16.065 | 8.39383 | 40.43 | 0.02751 | 0.01044 |
| | **$MgCl_2$** | 0.07724 | 3.57976 | 14.06563 | 8.14519 | 36.60 | 0.0294 | 0.01188 |
| **q = +56** | **KCl** | 0.09886 | 3.89898 | 17.83693 | 10.07563 | 37.17 | 0.0275 | 0.0094 |
| | **NaCl** | 0.10035 | 3.98535 | 18.4576 | 10.21093 | 37.85 | 0.02768 | 0.00909 |
| | **$CaCl_2$** | 0.10198 | 4.09401 | 18.99653 | 10.61264 | 37.42 | 0.02665 | 0.00884 |
| | **$MgCl_2$** | 0.06058 | 3.4492 | 18.57953 | 10.65996 | 36.00 | 0.02988 | 0.00913 |
| **q = +84** | **KCl** | 0.0833 | 3.85082 | 29.19301 | 14.01016 | 41.08 | 0.03287 | 0.00573 |
| | **NaCl** | 0.10213 | 4.02185 | 30.79973 | 14.5868 | 41.45 | 0.03183 | 0.00542 |
| | **$CaCl_2$** | 0.1043 | 4.01204 | 28.93855 | 14.16938 | 40.63 | 0.03089 | 0.00577 |
| | **$MgCl_2$** | 0.07845 | 3.66457 | 28.57006 | 14.40992 | 39.54 | 0.03251 | 0.0059 |

**Table S.7.** Same as Table S.6, but for the case of DND–$NH_2$.

| **DND's Charge** | **Dissolved Salt** | $\tau_{in}^{dip}$ / ps | $\tau_{c}^{dip}$ / ps | $\tau_{m}^{dip}$ / ps | $\tau_{corr}^{dip}$ / ps | $\theta_{tot}^{dip}$ / ° | $d_{c}^{dip}$ / $ps^{-1}$ | $d_{m}^{dip}$ / $ps^{-1}$ |
|---|---|---|---|---|---|---|---|---|
| **q = 0** | **KCl** | 0.09803 | 3.82421 | 13.53894 | 7.2417 | 39.76 | 0.03131 | 0.01238 |
| | **NaCl** | 0.10015 | 4.00659 | 15.435 | 7.96411 | 40.67 | 0.03116 | 0.01086 |
| | **$CaCl_2$** | 0.10065 | 3.88216 | 15.48775 | 8.07627 | 40.16 | 0.03142 | 0.01083 |
| | **$MgCl_2$** | 0.09974 | 3.98522 | 15.19604 | 7.88507 | 40.59 | 0.03112 | 0.01099 |
| **q = +28** | **KCl** | 0.08331 | 3.86859 | 31.98036 | 14.88264 | 41.61 | 0.03348 | 0.00525 |
| | **NaCl** | 0.10222 | 4.09615 | 34.64087 | 15.3588 | 42.93 | 0.03308 | 0.00483 |
| | **$CaCl_2$** | 0.10453 | 4.01266 | 32.37021 | 14.63898 | 42.34 | 0.03301 | 0.00521 |
| | **$MgCl_2$** | 0.08194 | 3.89342 | 33.47322 | 15.56957 | 41.64 | 0.03327 | 0.00502 |
| **q = +56** | **KCl** | 0.0751 | 3.49801 | 78.85178 | 41.216 | 37.22 | 0.03066 | 0.00214 |
| | **NaCl** | 0.08006 | 3.54086 | 88.90453 | 44.52133 | 37.97 | 0.03148 | 0.00207 |
| | **$CaCl_2$** | 0.05796 | 3.475 | 82.16318 | 43.40347 | 36.92 | 0.03044 | 0.00207 |
| | **$MgCl_2$** | 0.0948 | 3.45173 | 81.15085 | 42.16081 | 37.38 | 0.03127 | 0.00207 |
| **q = +84** | **KCl** | 0.07861 | 3.2298 | 331.2225 | 218.31221 | 29.70 | 0.02225 | 0.00052 |
| | **NaCl** | 0.07866 | 3.25566 | 413.62591 | 270.05705 | 29.83 | 0.02232 | 0.00049 |
| | **$CaCl_2$** | 0.06179 | 3.11535 | 377.14295 | 247.59176 | 29.78 | 0.02323 | 0.00048 |
| | **$MgCl_2$** | 0.06224 | 3.04579 | 368.90627 | 241.75343 | 29.74 | 0.02383 | 0.0005 |

**Table S.8.** Same as Table S.6, but for the case of DND–COOH.

| **DND's Charge** | **Dissolved Salt** | $\tau_{in}^{dip}$ / ps | $\tau_{c}^{dip}$ / ps | $\tau_{m}^{dip}$ / ps | $\tau_{corr}^{dip}$ / ps | $\theta_{tot}^{dip}$ / ° | $d_{c}^{dip}$ / $ps^{-1}$ | $d_{m}^{dip}$ / $ps^{-1}$ |
|---|---|---|---|---|---|---|---|---|
| **q = 0** | **KCl** | 0.0848 | 3.82421 | 22.39306 | 8.62504 | 47.06 | 0.04443 | 0.00756 |
| | **NaCl** | 0.08127 | 4.00659 | 20.06443 | 7.95922 | 46.73 | 0.0456 | 0.00833 |
| | **$CaCl_2$** | 0.08073 | 3.88216 | 18.87747 | 8.06852 | 44.82 | 0.04384 | 0.00888 |
| | **$MgCl_2$** | 0.10074 | 3.98522 | 20.22526 | 7.95827 | 47.01 | 0.04391 | 0.00828 |
| **q = –28** | **KCl** | 0.10496 | 3.86859 | 35.51699 | 14.11705 | 44.98 | 0.04012 | 0.00487 |
| | **NaCl** | 0.10283 | 4.09615 | 30.75229 | 12.39242 | 45.15 | 0.04064 | 0.0055 |
| | **$CaCl_2$** | 0.09758 | 4.01266 | 45.49722 | 19.87256 | 42.48 | 0.03851 | 0.00368 |
| | **$MgCl_2$** | 0.09316 | 3.89342 | 51.12286 | 23.88304 | 40.53 | 0.03525 | 0.00336 |
| **q = –56** | **KCl** | 0.09957 | 3.49801 | 44.89813 | 21.65666 | 40.00 | 0.03351 | 0.00373 |
| | **NaCl** | 0.10076 | 3.54086 | 43.15506 | 19.56906 | 41.61 | 0.03529 | 0.0039 |

| | CaCl2 | 0.08986 | 3.475 | 77.60349 | 40.90907 | 37.06 | 0.02933 | 0.00217 |
|---|---|---|---|---|---|---|---|---|
| | MgCl2 | 0.05823 | 3.45173 | 88.38432 | 50.54925 | 34.48 | 0.02579 | 0.00196 |
| **q = –84** | **KCl** | 0.08581 | 3.2298 | 77.15366 | 43.78132 | 34.88 | 0.02739 | 0.0022 |
| | **NaCl** | 0.07846 | 3.25566 | 60.41896 | 30.98359 | 37.87 | 0.03163 | 0.00288 |
| | **$CaCl_2$** | 0.07862 | 3.11535 | 140.95671 | 89.58144 | 31.01 | 0.02495 | 0.00124 |
| | **$MgCl_2$** | 0.07809 | 3.04579 | 164.48044 | 111.97313 | 28.74 | 0.02023 | 0.00102 |

**Table S.9.** Same as Table S.6, but for the case of DND–OH.

| DND's Charge | Dissolved Salt | $\tau_{in}^{dip}$ / ps | $\tau_{c}^{dip}$ / ps | $\tau_{m}^{dip}$ / ps | $\tau_{corr}^{dip}$ / ps | $\theta_{tot}^{dip}$ / ° | $d_{c}^{dip}$ / ps$^{-1}$ | $d_{m}^{dip}$ / ps$^{-1}$ |
|---|---|---|---|---|---|---|---|---|
| **q = 0** | **KCl** | 0.09979 | 4.13937 | 12.67506 | 5.95401 | 44.87 | 0.03504 | 0.0132 |
| | **NaCl** | 0.09902 | 4.1569 | 12.61647 | 5.98507 | 44.66 | 0.03472 | 0.01322 |
| | **$CaCl_2$** | 0.0982 | 4.10773 | 13.58546 | 6.2971 | 44.82 | 0.03525 | 0.01239 |
| | **$MgCl_2$** | 0.07922 | 4.04983 | 12.83891 | 5.89379 | 45.54 | 0.03666 | 0.01303 |

## S.2.1.2. Water's OH reorientation parameters

**Table S.10.** The EWIC parameters for the OH reorientations of water in the first hydration layer of DND–H with various surface chemistries.

| DND's Charge | Dissolved Salt | $\tau_{in}^{oh}$ / ps | $\tau_{c}^{oh}$ / ps | $\tau_{m}^{oh}$ / ps | $\tau_{corr}^{oh}$ / ps | $\theta_{tot}^{oh}$ / ° | $d_{c}^{oh}$ / ps$^{-1}$ | $d_{m}^{oh}$ / ps$^{-1}$ |
|---|---|---|---|---|---|---|---|---|
| **q = 0** | **KCl** | 0.12142 | 3.11636 | 15.37138 | 3.19053 | 64.32 | 0.07364 | 0.01089 |
| | **NaCl** | 0.1201 | 3.03949 | 18.24148 | 3.32216 | 65.20 | 0.07629 | 0.00939 |
| | **$CaCl_2$** | 0.12391 | 3.15439 | 16.08699 | 3.18378 | 64.92 | 0.07325 | 0.0106 |
| | **$MgCl_2$** | 0.12711 | 3.11379 | 20.01955 | 3.33645 | 66.54 | 0.07573 | 0.00861 |
| **q = +28** | **KCl** | 0.10575 | 4.71667 | 11.24145 | 5.91147 | 42.65 | 0.02846 | 0.01498 |
| | **NaCl** | 0.10829 | 4.97156 | 12.01114 | 6.07143 | 43.90 | 0.0281 | 0.01411 |
| | **$CaCl_2$** | 0.10084 | 4.48225 | 11.42689 | 6.14744 | 41.33 | 0.02836 | 0.01475 |
| | **$MgCl_2$** | 0.082 | 4.29991 | 11.21921 | 6.18347 | 40.52 | 0.02874 | 0.01488 |
| **q = +56** | **KCl** | 0.10062 | 4.05743 | 12.94021 | 7.58367 | 37.12 | 0.02636 | 0.01307 |
| | **NaCl** | 0.08448 | 4.29602 | 12.79358 | 7.44502 | 37.91 | 0.02585 | 0.01313 |
| | **$CaCl_2$** | 0.10051 | 4.08629 | 12.95827 | 7.96991 | 35.53 | 0.02427 | 0.01289 |
| | **$MgCl_2$** | 0.10107 | 4.21205 | 14.05662 | 8.28868 | 36.95 | 0.02523 | 0.01195 |
| **q = +84** | **KCl** | 0.11139 | 4.38285 | 16.18658 | 7.74503 | 43.50 | 0.03151 | 0.01061 |
| | **NaCl** | 0.0887 | 4.21055 | 15.13697 | 7.72799 | 41.93 | 0.03105 | 0.01104 |
| | **$CaCl_2$** | 0.10479 | 4.11304 | 15.89739 | 8.21487 | 41.11 | 0.03072 | 0.01054 |
| | **$MgCl_2$** | 0.10641 | 4.26319 | 16.26784 | 8.42471 | 40.89 | 0.02948 | 0.01045 |

**Table S.11.** Same as Table S.10, but for the case of DND–$NH_2$.

| DND's Charge | Dissolved Salt | $\tau_{in}^{oh}$ / ps | $\tau_{c}^{oh}$ / ps | $\tau_{m}^{oh}$ / ps | $\tau_{corr}^{oh}$ / ps | $\theta_{tot}^{oh}$ / ° | $d_{c}^{oh}$ / ps$^{-1}$ | $d_{m}^{oh}$ / ps$^{-1}$ |
|---|---|---|---|---|---|---|---|---|
| **q = 0** | **KCl** | 0.06099 | 3.52571 | 11.44787 | 6.51489 | 38.40 | 0.03211 | 0.01458 |
| | **NaCl** | 0.10326 | 4.08567 | 12.78403 | 6.86232 | 40.56 | 0.03025 | 0.0131 |
| | $CaCl_2$ | 0.10136 | 3.98114 | 13.35112 | 7.09269 | 40.62 | 0.0312 | 0.01255 |
| | $MgCl_2$ | 0.10457 | 4.2332 | 12.95901 | 6.86546 | 41.10 | 0.02985 | 0.01293 |
| q = +28 | KCl | 0.0857 | 3.83243 | 14.00856 | 7.02776 | 42.13 | 0.03452 | 0.01201 |
| | NaCl | 0.11085 | 4.23908 | 15.50304 | 7.03958 | 45.30 | 0.0348 | 0.01087 |
| | $CaCl_2$ | 0.11049 | 4.26547 | 14.75362 | 6.86632 | 44.71 | 0.03382 | 0.01146 |
| | $MgCl_2$ | 0.10784 | 4.05986 | 15.1 | 7.23756 | 43.46 | 0.03399 | 0.01117 |
| q = +56 | KCl | 0.11217 | 3.92972 | 18.36015 | 8.44096 | 43.80 | 0.03556 | 0.00914 |
| | NaCl | 0.10896 | 3.8354 | 17.855 | 8.11723 | 43.94 | 0.03657 | 0.00953 |
| | $CaCl_2$ | 0.10863 | 3.9424 | 18.29714 | 8.47359 | 43.60 | 0.03519 | 0.0092 |
| | $MgCl_2$ | 0.11672 | 4.03992 | 19.01531 | 8.52197 | 44.46 | 0.0355 | 0.00885 |
| q = +84 | KCl | 0.10504 | 4.03736 | 29.75869 | 15.24394 | 39.34 | 0.02909 | 0.00563 |
| | NaCl | 0.08581 | 4.03548 | 31.19988 | 15.76293 | 39.64 | 0.0297 | 0.00536 |
| | $CaCl_2$ | 0.10229 | 3.92536 | 28.07036 | 14.0723 | 39.79 | 0.03049 | 0.00609 |
| | $MgCl_2$ | 0.10577 | 4.04257 | 28.04591 | 14.27144 | 39.58 | 0.02942 | 0.006 |

**Table S.12.** Same as Table S.10, but for the case of DND–COOH.

| DND's Charge | Dissolved Salt | $\tau_{in}^{oh}$ / ps | $\tau_{c}^{oh}$ / ps | $\tau_{m}^{oh}$ / ps | $\tau_{corr}^{oh}$ / ps | $\theta_{tot}^{oh}$ / ° | $d_{c}^{oh}$ / ps$^{-1}$ | $d_{m}^{oh}$ / ps$^{-1}$ |
|---|---|---|---|---|---|---|---|---|
| q = 0 | KCl | 0.10894 | 3.25256 | 12.64887 | 4.48624 | 51.89 | 0.05509 | 0.01333 |
| | NaCl | 0.11073 | 3.29675 | 12.25895 | 4.23531 | 53.01 | 0.05599 | 0.01367 |
| | $CaCl_2$ | 0.11541 | 3.4871 | 12.46019 | 4.31982 | 53.14 | 0.05305 | 0.01346 |
| | $MgCl_2$ | 0.1136 | 3.41655 | 12.49947 | 4.30965 | 53.10 | 0.05412 | 0.01344 |
| q = –28 | KCl | 0.11545 | 3.50032 | 20.04581 | 7.401 | 48.86 | 0.04706 | 0.00833 |
| | NaCl | 0.1148 | 3.45982 | 16.96733 | 6.17414 | 49.88 | 0.04907 | 0.00986 |
| | $CaCl_2$ | 0.11005 | 3.40945 | 18.14813 | 6.67398 | 49.28 | 0.04897 | 0.00922 |
| | $MgCl_2$ | 0.10978 | 3.47664 | 26.68515 | 10.69415 | 45.83 | 0.04319 | 0.00626 |
| q = –56 | KCl | 0.11415 | 3.99713 | 32.73922 | 14.16831 | 43.57 | 0.03482 | 0.00516 |
| | NaCl | 0.11057 | 3.64795 | 25.13019 | 10.57239 | 44.77 | 0.03972 | 0.0067 |
| | $CaCl_2$ | 0.10822 | 3.62789 | 29.98238 | 12.91461 | 43.71 | 0.03842 | 0.00563 |
| | $MgCl_2$ | 0.0982 | 3.67278 | 43.12943 | 22.6804 | 37.76 | 0.02989 | 0.00391 |
| q = –84 | KCl | 0.07335 | 3.42389 | 47.1118 | 26.72868 | 35.36 | 0.02863 | 0.00355 |
| | NaCl | 0.10961 | 3.65285 | 37.01562 | 17.04516 | 41.62 | 0.03531 | 0.00457 |
| | $CaCl_2$ | 0.09511 | 3.49052 | 42.96813 | 22.48444 | 37.86 | 0.03156 | 0.00393 |
| | $MgCl_2$ | 0.0733 | 3.54145 | 58.97455 | 35.07482 | 33.49 | 0.02531 | 0.0029 |

**Table S.13.** Same as Table S.10, but for the case of DND–OH.

| DND's Charge | Dissolved Salt | $\tau_{in}^{oh}$ / ps | $\tau_{c}^{oh}$ / ps | $\tau_{m}^{oh}$ / ps | $\tau_{corr}^{oh}$ / ps | $\theta_{tot}^{oh}$ / ° | $d_{c}^{oh}$ / ps$^{-1}$ | $d_{m}^{oh}$ / ps$^{-1}$ |
|---|---|---|---|---|---|---|---|---|
| q = 0 | KCl | 0.11006 | 4.38607 | 11.18679 | 5.18982 | 46.84 | 0.03525 | 0.01498 |
| | NaCl | 0.08085 | 3.87792 | 10.22838 | 5.15341 | 43.81 | 0.03625 | 0.01633 |
| | $CaCl_2$ | 0.10231 | 4.16797 | 11.11882 | 5.26219 | 45.86 | 0.03599 | 0.0151 |
| | $MgCl_2$ | 0.09713 | 3.79553 | 9.92295 | 4.97572 | 43.94 | 0.03697 | 0.0169 |

## S.2.2. **Second hydration layer of DNDs**

### S.2.2.1. Water's dipole reorientation parameters

**Table S.14.** The EWIC parameters for the dipole reorientations of water in the second hydration layer of DND–H with various surface chemistries.

| DND's Charge | Dissolved Salt | $\tau_{in}^{dip}$ / ps | $\tau_{c}^{dip}$ / ps | $\tau_{m}^{dip}$ / ps | $\tau_{corr}^{dip}$ / ps | $\theta_{tot}^{dip}$ / ° | $d_{c}^{dip}$ / ps$^{-1}$ | $d_{m}^{dip}$ / ps$^{-1}$ |
|---|---|---|---|---|---|---|---|---|
| q = 0 | KCl | 0.11166 | 2.69776 | 55.91193 | 4.53673 | 70.83 | 0.09131 | 0.00303 |
| | NaCl | 0.11181 | 2.69807 | 70.18191 | 5.05595 | 71.12 | 0.09152 | 0.00258 |
| | $CaCl_2$ | 0.11766 | 2.70751 | 185.57519 | 10.16271 | 71.40 | 0.09144 | 0.00099 |
| | $MgCl_2$ | 0.11428 | 2.77626 | 55.82115 | 5.21227 | 68.91 | 0.08724 | 0.00318 |
| q = +28 | KCl | 0.11659 | 3.03506 | 38.75697 | 4.49298 | 67.95 | 0.07898 | 0.0046 |
| | NaCl | 0.11625 | 3.06894 | 31.97003 | 4.25949 | 66.97 | 0.07724 | 0.00551 |
| | $CaCl_2$ | 0.11491 | 3.04128 | 42.56867 | 4.73459 | 66.89 | 0.0778 | 0.00555 |
| | $MgCl_2$ | 0.12195 | 3.0783 | 56.03799 | 5.37553 | 68.08 | 0.07789 | 0.00395 |
| q = +56 | KCl | 0.11639 | 3.2741 | 21.53006 | 4.14283 | 63.35 | 0.06904 | 0.00793 |
| | NaCl | 0.11668 | 3.29336 | 21.38162 | 4.13761 | 63.06 | 0.06832 | 0.0082 |
| | $CaCl_2$ | 0.09577 | 3.19266 | 24.5558 | 4.2908 | 63.99 | 0.07139 | 0.00714 |
| | $MgCl_2$ | 0.11576 | 3.31228 | 21.50806 | 4.11936 | 63.61 | 0.06852 | 0.00793 |
| q = +84 | KCl | 0.09333 | 3.47179 | 19.09064 | 4.67155 | 59.27 | 0.06084 | 0.00893 |
| | NaCl | 0.11789 | 3.52715 | 19.97241 | 4.6794 | 59.54 | 0.06003 | 0.00892 |
| | $CaCl_2$ | 0.11972 | 3.48526 | 22.06182 | 4.53486 | 62.13 | 0.06367 | 0.00775 |
| | $MgCl_2$ | 0.09026 | 3.4093 | 18.21596 | 4.29187 | 60.11 | 0.06281 | 0.00965 |

**Table S.15.** Same as Table S.14, but for the case of DND–$NH_2$.

| DND's Charge | Dissolved Salt | $\tau_{in}^{dip}$ / ps | $\tau_{c}^{dip}$ / ps | $\tau_{m}^{dip}$ / ps | $\tau_{corr}^{dip}$ / ps | $\theta_{tot}^{dip}$ / ° | $d_{c}^{dip}$ / ps$^{-1}$ | $d_{m}^{dip}$ / ps$^{-1}$ |
|---|---|---|---|---|---|---|---|---|
| q = 0 | KCl | 0.11553 | 3.24452 | 22.7759 | 3.65373 | 66.82 | 0.07299 | 0.00756 |
| | NaCl | 0.09292 | 3.1725 | 25.81392 | 3.84306 | 66.89 | 0.07462 | 0.00685 |
| | $CaCl_2$ | 0.11499 | 3.2577 | 23.73944 | 3.98874 | 65.32 | 0.07131 | 0.00733 |
| | $MgCl_2$ | 0.11997 | 3.31055 | 26.34426 | 4.05523 | 65.92 | 0.07065 | 0.00708 |
| q = +28 | KCl | 0.1168 | 3.37811 | 19.28684 | 4.18067 | 61.58 | 0.06503 | 0.00908 |
| | NaCl | 0.11896 | 3.3954 | 20.15383 | 4.27448 | 61.46 | 0.06447 | 0.0089 |
| | $CaCl_2$ | 0.11027 | 3.27678 | 17.61771 | 4.02729 | 60.87 | 0.0662 | 0.00985 |
| | $MgCl_2$ | 0.11261 | 3.36678 | 17.7135 | 4.0532 | 61.15 | 0.06479 | 0.0097 |
| q = +56 | KCl | 0.11666 | 3.49294 | 20.21763 | 5.1975 | 57.75 | 0.05872 | 0.0083 |
| | NaCl | 0.1117 | 3.3706 | 19.71152 | 5.13757 | 57.20 | 0.06015 | 0.00871 |
| | $CaCl_2$ | 0.11369 | 3.42455 | 17.00113 | 4.72925 | 56.95 | 0.05895 | 0.00988 |
| | $MgCl_2$ | 0.11491 | 3.42383 | 21.13236 | 5.06547 | 58.99 | 0.06139 | 0.00798 |
| q = +84 | KCl | 0.1152 | 3.55162 | 22.5721 | 7.27609 | 51.78 | 0.05032 | 0.00744 |
| | NaCl | 0.11335 | 3.54414 | 20.72253 | 6.77191 | 51.89 | 0.0506 | 0.00807 |
| | $CaCl_2$ | 0.11756 | 3.59272 | 21.17412 | 6.65114 | 52.66 | 0.05085 | 0.00795 |

| | MgCl₂ | 0.11527 | 3.57153 | 20.57189 | 6.37947 | 53.20 | 0.05188 | 0.00816 |
|---|---|---|---|---|---|---|---|---|

**Table S.16.** Same as Table S.14, but for the case of DND–COOH.

| DND's Charge | Dissolved Salt | $\tau_{in}^{dip}$ / ps | $\tau_{c}^{dip}$ / ps | $\tau_{m}^{dip}$ / ps | $\tau_{corr}^{dip}$ / ps | $\theta_{tot}^{dip}$ / ° | $d_{c}^{dip}$ / ps$^{-1}$ | $d_{m}^{dip}$ / ps$^{-1}$ |
|---|---|---|---|---|---|---|---|---|
| q = 0 | KCl | 0.09194 | 3.26783 | 16.81392 | 3.42549 | 64.44 | 0.07019 | 0.01029 |
| | NaCl | 0.11381 | 3.2675 | 16.67022 | 3.31789 | 65.48 | 0.07133 | 0.01006 |
| | $CaCl_2$ | 0.11976 | 3.36206 | 20.09611 | 3.62296 | 65.80 | 0.06951 | 0.00861 |
| | $MgCl_2$ | 0.11407 | 3.35238 | 16.75072 | 3.42015 | 64.24 | 0.06826 | 0.0108 |
| q = –28 | KCl | 0.09455 | 3.27842 | 14.13868 | 3.33623 | 62.29 | 0.06768 | 0.0124 |
| | NaCl | 0.11442 | 3.30771 | 15.18724 | 3.50863 | 62.53 | 0.06748 | 0.01111 |
| | $CaCl_2$ | 0.11258 | 3.26397 | 22.42847 | 4.71541 | 60.85 | 0.06646 | 0.00771 |
| | $MgCl_2$ | 0.1201 | 3.42888 | 35.70922 | 7.94023 | 57.62 | 0.05967 | 0.00478 |
| q = –56 | KCl | 0.11517 | 3.3366 | 13.3467 | 3.55792 | 60.10 | 0.06435 | 0.01253 |
| | NaCl | 0.06963 | 3.09128 | 16.31849 | 3.91419 | 59.69 | 0.0687 | 0.01081 |
| | $CaCl_2$ | 0.11379 | 3.3617 | 23.32952 | 6.39689 | 55.08 | 0.05763 | 0.00733 |
| | $MgCl_2$ | 0.09227 | 3.35022 | 35.28489 | 10.33152 | 51.85 | 0.05341 | 0.005 |
| q = –84 | KCl | 0.11405 | 3.38643 | 13.90751 | 3.8241 | 56.73 | 0.06197 | 0.01205 |
| | NaCl | 0.11498 | 3.31587 | 15.98029 | 3.86546 | 57.42 | 0.0649 | 0.01087 |
| | $CaCl_2$ | 0.11545 | 3.41586 | 32.02352 | 9.14168 | 49.07 | 0.05383 | 0.00528 |
| | $MgCl_2$ | 0.11709 | 3.58314 | 45.13216 | 13.72688 | 48.33 | 0.04846 | 0.00386 |

**Table S.17.** Same as Table S.14, but for the case of DND–OH.

| DND's Charge | Dissolved Salt | $\tau_{in}^{dip}$ / ps | $\tau_{c}^{dip}$ / ps | $\tau_{m}^{dip}$ / ps | $\tau_{corr}^{dip}$ / ps | $\theta_{tot}^{dip}$ / ° | $d_{c}^{dip}$ / ps$^{-1}$ | $d_{m}^{dip}$ / ps$^{-1}$ |
|---|---|---|---|---|---|---|---|---|
| q = 0 | KCl | 0.11619 | 3.14839 | 25.16275 | 3.55633 | 68.42 | 0.07651 | 0.00686 |
| | NaCl | 0.11282 | 3.21427 | 19.62693 | 3.39522 | 66.77 | 0.07355 | 0.00868 |
| | $CaCl_2$ | 0.11214 | 3.22398 | 25.15029 | 3.98545 | 65.97 | 0.0727 | 0.00706 |
| | $MgCl_2$ | 0.09594 | 3.15611 | 34.31474 | 4.42988 | 65.84 | 0.07384 | 0.00688 |

## S.2.2.2. Water's OH reorientation parameters

**Table S.18.** The EWIC parameters for the OH reorientations of water in the second hydration layer of DND–H with various surface chemistries.

| DND's Charge | Dissolved Salt | $\tau_{in}^{oh}$ / ps | $\tau_{c}^{oh}$ / ps | $\tau_{m}^{oh}$ / ps | $\tau_{corr}^{oh}$ / ps | $\theta_{tot}^{oh}$ / ° | $d_{c}^{oh}$ / ps$^{-1}$ | $d_{m}^{oh}$ / ps$^{-1}$ |
|---|---|---|---|---|---|---|---|---|
| q = 0 | KCl | 0.12169 | 2.54351 | 136.15496 | 6.88057 | 72.79 | 0.09827 | 0.00126 |
| | NaCl | 0.11899 | 2.46407 | 70.15369 | 4.8824 | 71.49 | 0.10069 | 0.00245 |
| | $CaCl_2$ | 0.11833 | 2.52006 | 74.73188 | 5.31104 | 71.08 | 0.09795 | 0.00223 |
| | $MgCl_2$ | 0.12603 | 2.60936 | 163.37722 | 9.0606 | 71.54 | 0.09495 | 0.00108 |
| q = +28 | KCl | 0.12604 | 2.78015 | 52.00507 | 4.50472 | 70.60 | 0.08846 | 0.00323 |
| | NaCl | 0.12184 | 2.76796 | 59.58466 | 4.93594 | 70.20 | 0.08852 | 0.00306 |
| | $CaCl_2$ | 0.12801 | 2.80236 | 69.25819 | 5.11338 | 71.09 | 0.08811 | 0.0027 |
| | $MgCl_2$ | 0.12728 | 2.80126 | 65.20207 | 5.27144 | 69.96 | 0.08726 | 0.00313 |
| **q = +56** | **KCl** | 0.12615 | 3.0623 | 28.78902 | 3.94791 | 67.62 | 0.07796 | 0.00605 |
| | **NaCl** | 0.12518 | 3.07253 | 26.95326 | 3.80498 | 66.91 | 0.07715 | 0.00719 |
| | $CaCl_2$ | 0.1294 | 2.99448 | 32.39585 | 4.1205 | 67.75 | 0.07984 | 0.0056 |
| | $MgCl_2$ | 0.12318 | 3.03606 | 31.9619 | 4.07838 | 68.04 | 0.07907 | 0.00548 |
| **q = +84** | **KCl** | 0.12529 | 3.49355 | 20.27677 | 4.09844 | 63.95 | 0.06533 | 0.00829 |
| | **NaCl** | 0.11995 | 3.50863 | 16.84356 | 3.92253 | 61.96 | 0.06305 | 0.01032 |
| | $CaCl_2$ | 0.12525 | 3.45744 | 19.46282 | 3.85753 | 64.76 | 0.06671 | 0.00862 |
| | $MgCl_2$ | 0.12303 | 3.39793 | 23.4116 | 4.08345 | 65.29 | 0.06832 | 0.00759 |

**Table S.19.** Same as Table S.18, but for the case of DND–$NH_2$.

| DND's Charge | Dissolved Salt | $\tau_{in}^{oh}$ / ps | $\tau_{c}^{oh}$ / ps | $\tau_{m}^{oh}$ / ps | $\tau_{corr}^{oh}$ / ps | $\theta_{tot}^{oh}$ / ° | $d_{c}^{oh}$ / ps$^{-1}$ | $d_{m}^{oh}$ / ps$^{-1}$ |
|---|---|---|---|---|---|---|---|---|
| **q = 0** | **KCl** | 0.1256 | 2.97739 | 44.97235 | 4.15747 | 70.60 | 0.0826 | 0.00435 |
| | **NaCl** | 0.1252 | 2.95322 | 67.93402 | 4.97107 | 71.23 | 0.08373 | 0.00324 |
| | $CaCl_2$ | 0.12397 | 3.03407 | 36.78822 | 3.93837 | 69.81 | 0.0805 | 0.0051 |
| | $MgCl_2$ | 0.12396 | 2.99647 | 53.49448 | 4.63889 | 70.29 | 0.08187 | 0.00383 |
| **q = +28** | **KCl** | 0.12611 | 3.29119 | 23.74634 | 3.68855 | 67.35 | 0.07233 | 0.00757 |
| | **NaCl** | 0.12374 | 3.27334 | 20.44107 | 3.58883 | 66.44 | 0.07199 | 0.0084 |
| | $CaCl_2$ | 0.12244 | 3.23722 | 22.52328 | 3.705 | 66.08 | 0.07248 | 0.00842 |
| | $MgCl_2$ | 0.12343 | 3.31093 | 20.00539 | 3.61084 | 66.40 | 0.07117 | 0.00836 |
| **q = +56** | **KCl** | 0.11609 | 3.53709 | 13.93016 | 3.84488 | 59.77 | 0.06023 | 0.01206 |
| | **NaCl** | 0.12102 | 3.53613 | 15.38809 | 3.78538 | 61.86 | 0.06246 | 0.01098 |
| | $CaCl_2$ | 0.12183 | 3.42606 | 16.45334 | 3.71484 | 61.32 | 0.06343 | 0.01188 |
| | $MgCl_2$ | 0.11913 | 3.45071 | 16.54705 | 3.76459 | 62.71 | 0.06475 | 0.01046 |
| **q = +84** | **KCl** | 0.11869 | 3.79764 | 13.93567 | 4.68234 | 54.27 | 0.04995 | 0.01228 |
| | **NaCl** | 0.12002 | 3.74762 | 13.26919 | 4.6145 | 53.84 | 0.05024 | 0.01274 |
| | $CaCl_2$ | 0.11984 | 3.86104 | 13.379 | 4.41672 | 55.68 | 0.05086 | 0.01248 |
| | $MgCl_2$ | 0.11141 | 3.74014 | 12.68319 | 4.25642 | 55.43 | 0.05218 | 0.01321 |

**Table S.20.** Same as Table S.18, but for the case of DND–COOH.

| DND's Charge | Dissolved Salt | $\tau_{in}^{oh}$ / ps | $\tau_{c}^{oh}$ / ps | $\tau_{m}^{oh}$ / ps | $\tau_{corr}^{oh}$ / ps | $\theta_{tot}^{oh}$ / ° | $d_{c}^{oh}$ / ps$^{-1}$ | $d_{m}^{oh}$ / ps$^{-1}$ |
|---|---|---|---|---|---|---|---|---|
| **q = 0** | **KCl** | 0.12117 | 3.17328 | 21.37023 | 3.35495 | 68.41 | 0.07589 | 0.00805 |
| | **NaCl** | 0.11983 | 3.13419 | 21.62062 | 3.26128 | 69.18 | 0.07754 | 0.00788 |
| | $CaCl_2$ | 0.12445 | 3.18099 | 28.15779 | 3.55743 | 69.39 | 0.07645 | 0.00688 |
| | $MgCl_2$ | 0.12113 | 3.27945 | 19.81092 | 3.29788 | 68.10 | 0.07326 | 0.00896 |
| **q = –28** | **KCl** | 0.12233 | 3.17748 | 20.51327 | 3.26178 | 68.66 | 0.07604 | 0.00838 |
| | **NaCl** | 0.12157 | 3.16474 | 19.43331 | 3.27376 | 67.85 | 0.07567 | 0.00877 |
| | $CaCl_2$ | 0.11847 | 3.30643 | 18.31617 | 3.58878 | 64.84 | 0.06977 | 0.00963 |
| | $MgCl_2$ | 0.12685 | 3.60622 | 17.4081 | 4.22293 | 61.27 | 0.06071 | 0.00959 |
| **q = –56** | **KCl** | 0.1209 | 3.29344 | 13.10035 | 3.08562 | 64.55 | 0.06982 | 0.01279 |
| | **NaCl** | 0.12228 | 3.17842 | 17.60842 | 3.29891 | 66.25 | 0.07397 | 0.00977 |
| | $CaCl_2$ | 0.12148 | 3.5258 | 15.03365 | 3.85385 | 61.08 | 0.06189 | 0.01109 |
| | $MgCl_2$ | 0.1197 | 3.65213 | 14.48403 | 4.7125 | 54.96 | 0.05291 | 0.01159 |
| **q = –84** | **KCl** | 0.12031 | 3.46558 | 11.82026 | 3.2452 | 61.78 | 0.06364 | 0.01417 |
| | **NaCl** | 0.09493 | 3.19332 | 13.13413 | 3.13321 | 63.32 | 0.07058 | 0.01312 |
| | $CaCl_2$ | 0.12045 | 3.60725 | 15.54391 | 4.50789 | 57.23 | 0.05625 | 0.0109 |
| | $MgCl_2$ | 0.11406 | 3.6089 | 16.1861 | 5.59508 | 52.25 | 0.0501 | 0.01049 |

**Table S.21.** Same as Table S.18, but for the case of DND–OH.

| DND's Charge | Dissolved Salt | $\tau_{in}^{oh}$ / ps | $\tau_{c}^{oh}$ / ps | $\tau_{m}^{oh}$ / ps | $\tau_{corr}^{oh}$ / ps | $\theta_{tot}^{oh}$ / ° | $d_{c}^{oh}$ / ps$^{-1}$ | $d_{m}^{oh}$ / ps$^{-1}$ |
|---|---|---|---|---|---|---|---|---|
| **q = 0** | **KCl** | 0.12246 | 2.94594 | 37.54282 | 3.71859 | 71.10 | 0.08379 | 0.00475 |
| | **NaCl** | 0.12487 | 2.97815 | 38.57316 | 3.73137 | 71.28 | 0.08306 | 0.00478 |
| | $CaCl_2$ | 0.12189 | 3.07883 | 23.98606 | 3.49739 | 68.39 | 0.07817 | 0.00727 |
| | $MgCl_2$ | 0.12571 | 3.01141 | 54.38633 | 4.53826 | 70.19 | 0.08131 | 0.00455 |

## S.2.3. First hydration shell of ions (dissolved in aqueous solutions of the neutral DND–H)

### S.2.3.1. Water's dipole reorientation parameters

**Table S.22.** The EWIC parameters for the dipole reorientations of water in the first hydration shell of different ions that are dissolved in the aqueous solution of the neutral DND–H.

| Dissolved Ion | $\tau_{in}^{dip}$ / ps | $\tau_{c}^{dip}$ / ps | $\tau_{m}^{dip}$ / ps | $\tau_{corr}^{dip}$ / ps | $\theta_{tot}^{dip}$ / ° | $d_{c}^{dip}$ / ps$^{-1}$ | $d_{m}^{dip}$ / ps$^{-1}$ |
|---|---|---|---|---|---|---|---|
| $K^+$ | 0.0829 | 2.3039 | 7.42977 | 3.49639 | 44.71 | 0.06301 | 0.02249 |
| $Na^+$ | 0.09689 | 2.05955 | 13.18624 | 6.49247 | 40.60 | 0.06011 | 0.01264 |
| $Ca^{2+}$ | 0.05882 | 1.18104 | 20.78416 | 16.82372 | 21.55 | 0.03372 | 0.00802 |
| $Mg^{2+}$ | 0.07922 | 1.43321 | 39.42366 | 34.28938 | 17.47 | 0.01858 | 0.00423 |
| $Cl^-$ (KCl) | 0.10111 | 4.30518 | 6.46691 | 3.31861 | 47.73 | 0.03701 | 0.02583 |
| $Cl^-$ (NaCl) | 0.1012 | 4.13089 | 6.52057 | 3.35873 | 46.67 | 0.037 | 0.02595 |
| $Cl^-$ ($CaCl_2$) | 0.08199 | 4.05019 | 7.43666 | 3.48517 | 49.18 | 0.04087 | 0.02307 |
| $Cl^-$ ($MgCl_2$) | 0.08274 | 4.11281 | 8.06629 | 3.66594 | 50.41 | 0.04193 | 0.02069 |

### S.2.3.2. Water's OH reorientation parameters

**Table S.23.** Same as Table **S.22**, but for the case of OH reorientations.

| Dissolved Ion | $\tau_{in}^{oh}$ / ps | $\tau_{c}^{oh}$ / ps | $\tau_{m}^{oh}$ / ps | $\tau_{corr}^{oh}$ / ps | $\theta_{tot}^{oh}$ / ° | $d_{c}^{oh}$ / ps$^{-1}$ | $d_{m}^{oh}$ / ps$^{-1}$ |
|---|---|---|---|---|---|---|---|
| $K^+$ | 0.08561 | 2.6476 | 4.37994 | 2.30627 | 45.67 | 0.05758 | 0.03858 |
| $Na^+$ | 0.11331 | 2.9056 | 6.71265 | 2.74688 | 52.27 | 0.06225 | 0.02491 |
| $Ca^{2+}$ | 0.08728 | 4.02615 | 10.40717 | 4.61461 | 49.23 | 0.04135 | 0.01611 |
| $Mg^{2+}$ | 0.09371 | 5.94729 | 13.9436 | 8.73747 | 36.79 | 0.01763 | 0.01198 |
| $Cl^-$ (KCl) | 0.07496 | 2.19878 | 5.25782 | 3.9301 | 27.26 | 0.03213 | 0.03186 |
| $Cl^-$ (NaCl) | 0.06072 | 2.42466 | 5.33355 | 3.95745 | 27.98 | 0.03627 | 0.03151 |
| **$Cl^-$ ($CaCl_2$)** | 0.07567 | 2.74527 | 5.1646 | 3.87307 | 27.65 | 0.02497 | 0.03234 |
| **$Cl^-$ ($MgCl_2$)** | 0.06571 | 3.102 | 5.49449 | 3.97053 | 29.89 | 0.02774 | 0.03066 |

## S.2.4. **First hydration shell of Cations (dissolved in aqueous solutions of different DNDs)**

The results of this sub-section differ from those of Section B.2.3 in the sense that the EWIC parameters here are presented for water in the first hydration shell of cations dissolved in the aqueous solutions of various DNDs with a distinct surface chemistry. As a reminder, only one DND particle is solvated in the solution.

### S.2.4.1. Water's dipole reorientation parameters

**Table S.24.** The EWIC parameters for the dipole reorientations of water in the first hydration shell of different cations that are dissolved in the aqueous solutions of DND–H with various surface chemistries.

| DND's Charge | Dissolved Cation | $\tau_{in}^{dip}$ / ps | $\tau_{c}^{dip}$ / ps | $\tau_{m}^{dip}$ / ps | $\tau_{corr}^{dip}$ / ps | $\theta_{tot}^{dip}$ / ° | $d_{c}^{dip}$ / ps$^{-1}$ | $d_{m}^{dip}$ / ps$^{-1}$ |
|---|---|---|---|---|---|---|---|---|
| q = 0 | $K^+$ | 0.0829 | 2.3039 | 7.42977 | 3.49639 | 44.71 | 0.06301 | 0.02249 |
| | $Na^+$ | 0.09689 | 2.05955 | 13.18624 | 6.49247 | 40.60 | 0.06011 | 0.01264 |
| | $Ca^{2+}$ | 0.05882 | 1.18104 | 20.78416 | 16.82372 | 21.55 | 0.03372 | 0.00802 |
| | $Mg^{2+}$ | 0.07922 | 1.43321 | 39.42366 | 34.28938 | 17.47 | 0.01858 | 0.00423 |
| q = +28 | $K^+$ | 0.10374 | 2.40738 | 7.29096 | 3.47959 | 44.52 | 0.05954 | 0.02289 |
| | $Na^+$ | 0.07619 | 1.9397 | 12.63357 | 6.37409 | 39.87 | 0.06209 | 0.0132 |
| | $Ca^{2+}$ | 0.04724 | 1.12702 | 20.68639 | 16.73477 | 21.57 | 0.03631 | 0.00806 |
| | $Mg^{2+}$ | 0.07979 | 1.56517 | 39.20475 | 34.01707 | 17.62 | 0.01796 | 0.00426 |
| q = +56 | $K^+$ | 0.10508 | 2.47426 | 7.88289 | 3.5578 | 45.92 | 0.06082 | 0.02138 |
| | $Na^+$ | 0.09899 | 2.08028 | 12.92148 | 6.38736 | 40.54 | 0.05942 | 0.0129 |
| | $Ca^{2+}$ | 0.04551 | 1.01164 | 20.55899 | 16.69122 | 21.38 | 0.03877 | 0.00811 |
| | $Mg^{2+}$ | 0.08164 | 1.67836 | 40.72044 | 35.27044 | 17.73 | 0.01725 | 0.0041 |
| q = +84 | $K^+$ | 0.10437 | 2.47504 | 7.49595 | 3.5 | 45.27 | 0.05964 | 0.02226 |
| | $Na^+$ | 0.09817 | 2.07532 | 12.9712 | 6.38938 | 40.60 | 0.05979 | 0.01288 |
| | $Ca^{2+}$ | 0.04778 | 1.14513 | 21.07545 | 17.0369 | 21.61 | 0.03647 | 0.00791 |
| | $Mg^{2+}$ | 0.0766 | 1.3072 | 39.46178 | 34.33365 | 17.45 | 0.02031 | 0.00422 |

**Table S.25.** Same as Table S.24, but for the case of DND–$NH_2$.

| DND's Charge | Dissolved Cation | $\tau_{in}^{dip}$ / ps | $\tau_{c}^{dip}$ / ps | $\tau_{m}^{dip}$ / ps | $\tau_{corr}^{dip}$ / ps | $\theta_{tot}^{dip}$ / ° | $d_{c}^{dip}$ / ps$^{-1}$ | $d_{m}^{dip}$ / ps$^{-1}$ |
|---|---|---|---|---|---|---|---|---|
| q = 0 | $K^+$ | 0.10781 | 2.56877 | 8.28298 | 3.70633 | 46.25 | 0.05917 | 0.02018 |
| | $Na^+$ | 0.10087 | 2.19523 | 14.03397 | 6.81992 | 40.95 | 0.05721 | 0.01189 |
| | $Ca^{2+}$ | 0.05709 | 1.09998 | 20.86294 | 16.88738 | 21.53 | 0.03592 | 0.00799 |
| | $Mg^{2+}$ | 0.08061 | 1.69753 | 41.65024 | 36.04191 | 17.78 | 0.01692 | 0.00401 |
| q = +28 | $K^+$ | 0.1018 | 2.38184 | 7.52371 | 3.50642 | 45.12 | 0.06159 | 0.02219 |
| | $Na^+$ | 0.07964 | 2.01229 | 12.91599 | 6.39736 | 40.42 | 0.06151 | 0.01292 |
| | $Ca^{2+}$ | 0.05788 | 1.15195 | 20.79339 | 16.83256 | 21.54 | 0.03507 | 0.00802 |
| | $Mg^{2+}$ | 0.07846 | 1.4283 | 40.27604 | 34.96305 | 17.59 | 0.019 | 0.00414 |
| q = +56 | $K^+$ | 0.08983 | 2.54325 | 7.88044 | 3.53479 | 46.38 | 0.0605 | 0.0213 |
| | $Na^+$ | 0.09815 | 2.11124 | 13.1076 | 6.44505 | 40.69 | 0.0589 | 0.01273 |
| | $Ca^{2+}$ | 0.04523 | 0.97869 | 20.43708 | 16.59763 | 21.35 | 0.04109 | 0.00816 |
| | $Mg^{2+}$ | 0.08057 | 1.53559 | 39.44362 | 34.26602 | 17.55 | 0.01771 | 0.00423 |
| **q = +84** | $K^+$ | 0.1046 | 2.4729 | 7.5373 | 3.48915 | 45.50 | 0.05999 | 0.02216 |
| | $Na^+$ | 0.09965 | 2.11818 | 13.38417 | 6.51263 | 40.95 | 0.05956 | 0.01246 |
| | $Ca^{2+}$ | 0.05576 | 1.06158 | 20.66181 | 16.75036 | 21.45 | 0.03729 | 0.00807 |
| | $Mg^{2+}$ | 0.07908 | 1.43758 | 39.74972 | 34.54095 | 17.53 | 0.01918 | 0.0042 |

**Table S.26.** Same as Table S.24, but for the case of DND–COOH.

| DND's Charge | Dissolved Cation | $\tau_{in}^{dip}$ / ps | $\tau_{c}^{dip}$ / ps | $\tau_{m}^{dip}$ / ps | $\tau_{corr}^{dip}$ / ps | $\theta_{tot}^{dip}$ / ° | $d_{c}^{dip}$ / ps$^{-1}$ | $d_{m}^{dip}$ / ps$^{-1}$ |
|---|---|---|---|---|---|---|---|---|
| **q = 0** | $K^+$ | 0.12026 | 2.81633 | 11.28367 | 4.18538 | 50.42 | 0.06114 | 0.01504 |
| | $Na^+$ | 0.09876 | 2.13954 | 13.72261 | 6.70414 | 40.81 | 0.0584 | 0.01215 |
| | $Ca^{2+}$ | 0.05595 | 1.10087 | 21.20264 | 17.16102 | 21.54 | 0.03683 | 0.00786 |
| | $Mg^{2+}$ | 0.0797 | 1.62684 | 40.51071 | 35.07157 | 17.77 | 0.01764 | 0.00412 |
| **q = –28** | $K^+$ | 0.12337 | 2.86127 | 17.12025 | 5.67332 | 51.28 | 0.06166 | 0.00982 |
| | $Na^+$ | 0.10431 | 2.33108 | 15.90806 | 7.52812 | 41.52 | 0.05515 | 0.0105 |
| | $Ca^{2+}$ | 0.0644 | 1.59551 | 23.89434 | 19.05123 | 22.33 | 0.02663 | 0.00698 |
| | $Mg^{2+}$ | 0.08111 | 1.81068 | 45.13935 | 39.07959 | 17.75 | 0.01535 | 0.0037 |
| **q = –56** | $K^+$ | 0.10259 | 2.88231 | 26.00231 | 8.04311 | 51.28 | 0.06136 | 0.00646 |
| | $Na^+$ | 0.10479 | 2.39186 | 17.01082 | 8.01448 | 41.57 | 0.05377 | 0.0098 |
| | $Ca^{2+}$ | 0.07342 | 2.40611 | 27.62062 | 21.53628 | 23.49 | 0.01939 | 0.00604 |
| | $Mg^{2+}$ | 0.08171 | 2.18812 | 51.71268 | 44.50497 | 18.13 | 0.01329 | 0.00323 |
| **q = –84** | $K^+$ | 0.12602 | 2.93766 | 33.94203 | 10.61855 | 50.36 | 0.05854 | 0.00493 |
| | $Na^+$ | 0.08752 | 2.5616 | 22.12357 | 10.14302 | 41.90 | 0.05122 | 0.00757 |
| | $Ca^{2+}$ | 0.07173 | 2.39859 | 31.56679 | 24.64478 | 23.39 | 0.01951 | 0.00528 |
| | $Mg^{2+}$ | 0.08208 | 2.47971 | 56.26651 | 48.3798 | 18.18 | 0.01239 | 0.00297 |

**Table S.27.** Same as Table S.24, but for the case of DND–OH.

| DND's Charge | Dissolved Cation | $\tau_{in}^{dip}$ / ps | $\tau_{c}^{dip}$ / ps | $\tau_{m}^{dip}$ / ps | $\tau_{corr}^{dip}$ / ps | $\theta_{tot}^{dip}$ / ° | $d_{c}^{dip}$ / ps$^{-1}$ | $d_{m}^{dip}$ / ps$^{-1}$ |
|---|---|---|---|---|---|---|---|---|
| **q = 0** | $K^+$ | 0.10823 | 2.57418 | 8.55948 | 3.71305 | 47.08 | 0.06062 | 0.01952 |
| | $Na^+$ | 0.09595 | 2.08316 | 13.50571 | 6.63965 | 40.61 | 0.05954 | 0.01235 |
| | $Ca^{2+}$ | 0.03429 | 0.94298 | 20.90945 | 16.96591 | 21.39 | 0.04212 | 0.00797 |
| | $Mg^{2+}$ | 0.08414 | 1.87546 | 40.99642 | 35.48122 | 17.80 | 0.01478 | 0.00407 |

## S.2.4.2. Water's OH reorientation parameters

**Table S.28.** The EWIC parameters for the OH reorientations of water in the first hydration shell of different cations that are dissolved in the aqueous solutions of DND–H with various surface chemistries.

| DND's Charge | Dissolved Cation | $\tau_{in}^{oh}$ / ps | $\tau_{c}^{oh}$ / ps | $\tau_{m}^{oh}$ / ps | $\tau_{corr}^{oh}$ / ps | $\theta_{tot}^{oh}$ / ° | $d_{c}^{oh}$ / ps$^{-1}$ | $d_{m}^{oh}$ / ps$^{-1}$ |
|---|---|---|---|---|---|---|---|---|
| **q = 0** | $K^+$ | 0.08561 | 2.6476 | 4.37994 | 2.30627 | 45.67 | 0.05758 | 0.03858 |
| | $Na^+$ | 0.11331 | 2.9056 | 6.71265 | 2.74688 | 52.27 | 0.06225 | 0.02491 |
| | **$Ca^{2+}$** | 0.08728 | 4.02615 | 10.40717 | 4.61461 | 49.23 | 0.04135 | 0.01611 |
| | **$Mg^{2+}$** | 0.09371 | 5.94729 | 13.9436 | 8.73747 | 36.79 | 0.01763 | 0.01198 |
| **q = +28** | **$K^{+}$** | 0.10831 | 3.0176 | 4.49228 | 2.31145 | 47.88 | 0.05307 | 0.03718 |
| | **$Na^{+}$** | 0.11209 | 2.92592 | 6.43637 | 2.73114 | 51.43 | 0.0605 | 0.02595 |
| | **$Ca^{2+}$** | 0.10774 | 4.07014 | 10.09485 | 4.60654 | 48.53 | 0.04005 | 0.01656 |
| | **$Mg^{2+}$** | 0.09186 | 5.36433 | 12.96585 | 8.4724 | 34.71 | 0.01773 | 0.01289 |
| **q = +56** | **$K^{+}$** | 0.10652 | 2.99128 | 4.61418 | 2.3244 | 48.59 | 0.05479 | 0.03616 |
| | **$Na^{+}$** | 0.11231 | 2.91551 | 6.46461 | 2.70885 | 51.85 | 0.0614 | 0.02584 |
| | **$Ca^{2+}$** | 0.08671 | 3.97347 | 10.11479 | 4.63034 | 48.23 | 0.04074 | 0.01654 |
| | **$Mg^{2+}$** | 0.07034 | 5.64109 | 13.89368 | 8.73885 | 36.36 | 0.01826 | 0.01204 |
| **q = +84** | **$K^{+}$** | 0.10971 | 3.12665 | 4.86287 | 2.33356 | 50.70 | 0.05553 | 0.03437 |
| | **$Na^{+}$** | 0.1123 | 2.93538 | 6.75996 | 2.72377 | 52.94 | 0.06273 | 0.02471 |
| | **$Ca^{2+}$** | 0.1119 | 4.234 | 10.61605 | 4.65831 | 49.82 | 0.03999 | 0.01576 |
| | **$Mg^{2+}$** | 0.11455 | 5.85812 | 13.6981 | 8.68274 | 36.20 | 0.01743 | 0.01222 |

**Table S.29.** Same as Table S.28, but for the case of DND–$NH_2$.

| DND's Charge | Dissolved Cation | $\tau_{in}^{oh}$ / ps | $\tau_{c}^{oh}$ / ps | $\tau_{m}^{oh}$ / ps | $\tau_{corr}^{oh}$ / ps | $\theta_{tot}^{oh}$ / ° | $d_{c}^{oh}$ / ps$^{-1}$ | $d_{m}^{oh}$ / ps$^{-1}$ |
|---|---|---|---|---|---|---|---|---|
| **q = 0** | **$K^{+}$** | 0.11064 | 2.99682 | 5.37772 | 2.43437 | 51.22 | 0.05838 | 0.03152 |
| | **$Na^{+}$** | 0.11594 | 3.02245 | 7.41656 | 2.85511 | 53.93 | 0.06242 | 0.02249 |
| | **$Ca^{2+}$** | 0.10749 | 4.06692 | 10.56723 | 4.66731 | 49.34 | 0.04106 | 0.0158 |
| | **$Mg^{2+}$** | 0.11348 | 5.6313 | 13.71199 | 8.79784 | 35.48 | 0.01757 | 0.0122 |
| **q = +28** | **$K^{+}$** | 0.11071 | 3.11902 | 4.92157 | 2.30477 | 51.24 | 0.05598 | 0.03457 |
| | **$Na^{+}$** | 0.11467 | 2.96315 | 6.80025 | 2.73977 | 52.95 | 0.06213 | 0.02457 |
| | **$Ca^{2+}$** | 0.09093 | 4.06027 | 10.60568 | 4.6282 | 49.44 | 0.04116 | 0.016 |
| | **$Mg^{2+}$** | 0.09442 | 5.65596 | 13.29246 | 8.60366 | 35.32 | 0.01736 | 0.01255 |
| **q = +56** | **$K^{+}$** | 0.10461 | 2.8119 | 4.52582 | 2.31847 | 47.31 | 0.05585 | 0.03703 |
| | **$Na^{+}$** | 0.1151 | 3.00857 | 6.83476 | 2.74446 | 53.19 | 0.06154 | 0.02445 |
| | **$Ca^{2+}$** | 0.10774 | 4.05987 | 10.10527 | 4.60275 | 48.61 | 0.04026 | 0.01651 |
| | **$Mg^{2+}$** | 0.11649 | 5.67856 | 13.08744 | 8.53452 | 35.05 | 0.01701 | 0.01274 |
| **q = +84** | **$K^{+}$** | 0.10306 | 2.76817 | 4.39148 | 2.29673 | 46.21 | 0.05448 | 0.03848 |
| | **$Na^{+}$** | 0.09009 | 2.71314 | 6.54478 | 2.73721 | 51.10 | 0.06484 | 0.02566 |
| | **$Ca^{2+}$** | 0.08921 | 4.01065 | 10.52938 | 4.64781 | 49.19 | 0.04141 | 0.01596 |
| | **$Mg^{2+}$** | 0.1154 | 6.06067 | 13.98069 | 8.67218 | 37.19 | 0.01763 | 0.01199 |

**Table S.30.** Same as Table S.28, but for the case of DND–COOH.

| DND's Charge | Dissolved Cation | $\tau_{in}^{oh}$ / ps | $\tau_{c}^{oh}$ / ps | $\tau_{m}^{oh}$ / ps | $\tau_{corr}^{oh}$ / ps | $\theta_{tot}^{oh}$ / ° | $d_{c}^{oh}$ / ps$^{-1}$ | $d_{m}^{oh}$ / ps$^{-1}$ |
|---|---|---|---|---|---|---|---|---|
| **q = 0** | **$K^{+}$** | 0.11516 | 2.95992 | 6.49016 | 2.51658 | 54.76 | 0.06469 | 0.02652 |
| | **$Na^{+}$** | 0.11622 | 3.01206 | 7.18212 | 2.80619 | 53.60 | 0.06209 | 0.02327 |
| | **$Ca^{2+}$** | 0.10767 | 4.02656 | 10.33512 | 4.68678 | 48.47 | 0.04045 | 0.01617 |
| | **$Mg^{2+}$** | 0.1153 | 5.59997 | 13.81202 | 8.8038 | 35.73 | 0.01784 | 0.01209 |
| **q = –28** | **$K^{+}$** | 0.13503 | 2.95695 | 13.95883 | 3.39718 | 60.94 | 0.07353 | 0.0122 |
| | $Na^{+}$ | 0.11568 | 2.87198 | 8.43444 | 3.12732 | 53.20 | 0.06446 | 0.01991 |
| | $Ca^{2+}$ | 0.08666 | 3.88429 | 11.68916 | 5.09349 | 48.49 | 0.04196 | 0.01434 |
| | $Mg^{2+}$ | 0.11597 | 5.98671 | 16.47864 | 9.93664 | 37.45 | 0.01805 | 0.01018 |
| **q = –56** | $K^{+}$ | 0.13682 | 2.90394 | 19.73639 | 4.60185 | 58.98 | 0.07233 | 0.00856 |
| | $Na^{+}$ | 0.12316 | 3.0124 | 10.2809 | 3.42259 | 55.20 | 0.06448 | 0.01628 |
| | $Ca^{2+}$ | 0.11182 | 4.03976 | 13.75981 | 5.63476 | 49.49 | 0.04152 | 0.01219 |
| | $Mg^{2+}$ | 0.11633 | 6.02653 | 18.37792 | 10.99922 | 37.41 | 0.01791 | 0.00908 |
| **q = –84** | $K^{+}$ | 0.13628 | 2.8651 | 32.59284 | 7.22947 | 57.42 | 0.07115 | 0.00515 |
| | $Na^{+}$ | 0.12395 | 3.05397 | 13.85178 | 4.30218 | 54.85 | 0.0631 | 0.01207 |
| | $Ca^{2+}$ | 0.11001 | 4.05022 | 16.6068 | 6.68035 | 48.85 | 0.04065 | 0.01006 |
| | $Mg^{2+}$ | 0.11463 | 6.07132 | 20.78753 | 12.17721 | 37.65 | 0.01798 | 0.00805 |

**Table S.31.** Same as Table S.28, but for the case of DND–OH.

| DND's Charge | Dissolved Cation | $\tau_{in}^{oh}$ / ps | $\tau_{c}^{oh}$ / ps | $\tau_{m}^{oh}$ / ps | $\tau_{corr}^{oh}$ / ps | $\theta_{tot}^{oh}$ / ° | $d_{c}^{oh}$ / ps$^{-1}$ | $d_{m}^{oh}$ / ps$^{-1}$ |
|---|---|---|---|---|---|---|---|---|
| **q = 0** | $K^{+}$ | 0.11203 | 2.99982 | 5.07244 | 2.38171 | 50.10 | 0.05659 | 0.03353 |
| | $Na^{+}$ | 0.11029 | 2.85722 | 6.59047 | 2.78359 | 51.18 | 0.06165 | 0.02533 |
| | $Ca^{2+}$ | 0.1076 | 4.06457 | 10.39022 | 4.66934 | 48.87 | 0.04058 | 0.01606 |
| | $Mg^{2+}$ | 0.12017 | 6.24464 | 14.70689 | 8.9684 | 37.83 | 0.01759 | 0.01139 |

## S.2.5. First hydration shell of $Cl^{-}$ anion (dissolved in aqueous solutions of different DNDs)

### S.2.5.1. Water's dipole reorientation parameters

**Table S.32.** The EWIC parameters for the dipole reorientations of water in the first hydration shell of $Cl^{-}$ anions of different salts that are dissolved in the aqueous solutions of DND–H with various surface chemistries.

| DND's Charge | Dissolved Anion | $\tau_{in}^{dip}$ / ps | $\tau_{c}^{dip}$ / ps | $\tau_{m}^{dip}$ / ps | $\tau_{corr}^{dip}$ / ps | $\theta_{tot}^{dip}$ / ° | $d_{c}^{dip}$ / ps$^{-1}$ | $d_{m}^{dip}$ / ps$^{-1}$ |
|---|---|---|---|---|---|---|---|---|
| **q = 0** | $Cl^{-}$ (KCl) | 0.10111 | 4.30518 | 6.46691 | 3.31861 | 47.73 | 0.03701 | 0.02583 |
| | $Cl^{-}$ (NaCl) | 0.1012 | 4.13089 | 6.52057 | 3.35873 | 46.67 | 0.037 | 0.02595 |
| | $Cl^{-}$ ($CaCl_2$) | 0.08199 | 4.05019 | 7.43666 | 3.48517 | 49.18 | 0.04087 | 0.02307 |
| | $Cl^{-}$ ($MgCl_2$) | 0.08274 | 4.11281 | 8.06629 | 3.66594 | 50.41 | 0.04193 | 0.02069 |
| **q = +28** | $Cl^{-}$ (KCl) | 0.06195 | 3.73992 | 6.29004 | 3.37229 | 44.63 | 0.03894 | 0.02676 |
| | $Cl^{-}$ (NaCl) | 0.10432 | 4.3642 | 7.35205 | 3.46001 | 50.00 | 0.03874 | 0.02308 |
| | $Cl^{-}$ ($CaCl_2$) | 0.102 | 4.27495 | 7.45701 | 3.58841 | 49.01 | 0.0387 | 0.02242 |
| | $Cl^-$ ($MgCl_2$) | 0.10733 | 3.99629 | 12.9181 | 4.32059 | 56.03 | 0.04948 | 0.01297 |
| **q = +56** | $Cl^-$ (KCl) | 0.10436 | 4.17102 | 9.18871 | 3.9197 | 51.48 | 0.04243 | 0.01825 |
| | $Cl^-$ (NaCl) | 0.10564 | 4.159 | 9.09119 | 3.94701 | 50.88 | 0.04193 | 0.0184 |
| | $Cl^-$ ($CaCl_2$) | 0.10241 | 4.12893 | 9.07887 | 3.96554 | 50.69 | 0.04204 | 0.01837 |
| | $Cl^-$ ($MgCl_2$) | 0.08546 | 3.91666 | 12.00458 | 4.49783 | 52.94 | 0.04703 | 0.01393 |
| **q = +84** | $Cl^-$ (KCl) | 0.10467 | 4.06102 | 11.20408 | 4.41476 | 52.23 | 0.04454 | 0.0149 |
| | $Cl^-$ (NaCl) | 0.10694 | 4.07264 | 11.71443 | 4.53969 | 52.25 | 0.04438 | 0.01432 |
| | $Cl^-$ ($CaCl_2$) | 0.10687 | 4.17126 | 10.85512 | 4.23582 | 53.10 | 0.04431 | 0.01537 |
| | $Cl^-$ ($MgCl_2$) | 0.10925 | 4.03769 | 13.90305 | 4.76069 | 54.51 | 0.04733 | 0.0121 |

**Table S.33.** Same as Table S.32, but for the case of DND–$NH_2$.

| DND's Charge | Dissolved Anion | $\tau_{in}^{dip}$ / ps | $\tau_{c}^{dip}$ / ps | $\tau_{m}^{dip}$ / ps | $\tau_{corr}^{dip}$ / ps | $\theta_{tot}^{dip}$ / ° | $d_{c}^{dip}$ / ps$^{-1}$ | $d_{m}^{dip}$ / ps$^{-1}$ |
|---|---|---|---|---|---|---|---|---|
| **q = 0** | $Cl^-$ (KCl) | 0.10095 | 4.30464 | 6.47757 | 3.33038 | 47.61 | 0.03681 | 0.02582 |
| | $Cl^-$ (NaCl) | 0.10371 | 4.49074 | 7.14219 | 3.40639 | 50.48 | 0.03836 | 0.02346 |
| | $Cl^-$ ($CaCl_2$) | 0.10075 | 4.33328 | 7.19086 | 3.48013 | 49.33 | 0.03848 | 0.02329 |
| | $Cl^-$ ($MgCl_2$) | 0.10987 | 4.13048 | 12.25587 | 4.10686 | 56.58 | 0.04837 | 0.01391 |
| **q = +28** | $Cl^-$ (KCl) | 0.06435 | 3.80757 | 7.06922 | 3.51827 | 46.68 | 0.04049 | 0.02405 |
| | $Cl^-$ (NaCl) | 0.10195 | 4.17747 | 7.36069 | 3.54923 | 48.77 | 0.03924 | 0.02276 |
| | $Cl^-$ ($CaCl_2$) | 0.10535 | 4.28422 | 8.67251 | 3.71062 | 52.14 | 0.04198 | 0.01945 |
| | $Cl^-$ ($MgCl_2$) | 0.10532 | 4.11865 | 9.99276 | 3.9426 | 53.45 | 0.04525 | 0.01675 |
| **q = +56** | $Cl^-$ (KCl) | 0.10974 | 4.18946 | 12.73189 | 4.23439 | 56.85 | 0.04807 | 0.01317 |
| | $Cl^-$ (NaCl) | 0.1103 | 4.21056 | 12.56465 | 4.20909 | 56.21 | 0.04707 | 0.01378 |
| | $Cl^-$ ($CaCl_2$) | 0.10616 | 4.1493 | 11.05887 | 4.10435 | 54.49 | 0.04601 | 0.01521 |
| | $Cl^-$ ($MgCl_2$) | 0.10893 | 3.97426 | 15.84387 | 4.82973 | 56.79 | 0.05059 | 0.01054 |
| **q = +84** | $Cl^-$ | 0.09093 | 3.74782 | 27.82691 | 7.00459 | 56.82 | 0.05368 | 0.00614 |
| | (KCl) | | | | | | | |
| | Cl⁻ (NaCl) | 0.09095 | 3.74982 | 31.68966 | 7.73457 | 56.83 | 0.05367 | 0.00538 |
| | Cl⁻ ($CaCl_2$) | 0.11227 | 3.85677 | 23.57192 | 6.04767 | 57.69 | 0.05312 | 0.00712 |
| | Cl⁻ ($MgCl_2$) | 0.11359 | 3.79735 | 31.30362 | 7.36536 | 57.57 | 0.0538 | 0.0055 |

**Table S.34.** Same as Table S.32, but for the case of DND–COOH.

| DND's Charge | Dissolved Anion | $\tau_{in}^{dip}$ / ps | $\tau_{c}^{dip}$ / ps | $\tau_{m}^{dip}$ / ps | $\tau_{corr}^{dip}$ / ps | $\theta_{tot}^{dip}$ / ° | $d_{c}^{dip}$ / ps$^{-1}$ | $d_{m}^{dip}$ / ps$^{-1}$ |
|---|---|---|---|---|---|---|---|---|
| q = 0 | Cl⁻ (KCl) | 0.10287 | 4.4629 | 7.34205 | 3.34511 | 52.22 | 0.04042 | 0.02288 |
| | Cl⁻ (NaCl) | 0.10317 | 4.4915 | 6.79874 | 3.35595 | 49.47 | 0.03727 | 0.02462 |
| | Cl⁻ ($CaCl_2$) | 0.10426 | 4.45876 | 7.84523 | 3.53992 | 51.88 | 0.0402 | 0.02126 |
| | Cl⁻ ($MgCl_2$) | 0.10738 | 4.05563 | 12.66128 | 4.18742 | 56.72 | 0.04955 | 0.01331 |
| q = –28 | Cl⁻ (KCl) | 0.10427 | 4.32519 | 6.84537 | 3.33346 | 49.31 | 0.03812 | 0.02476 |
| | Cl⁻ (NaCl) | 0.08157 | 4.10978 | 7.22297 | 3.41528 | 49.59 | 0.04071 | 0.02341 |
| | Cl⁻ ($CaCl_2$) | 0.07835 | 4.02917 | 7.33266 | 3.52546 | 48.27 | 0.03991 | 0.02322 |
| | Cl⁻ ($MgCl_2$) | 0.10591 | 4.09227 | 10.52254 | 3.97368 | 54.46 | 0.04666 | 0.01588 |
| q = –56 | Cl⁻ (KCl) | 0.10165 | 4.3766 | 6.95712 | 3.31491 | 49.93 | 0.0384 | 0.02461 |
| | Cl⁻ (NaCl) | 0.10409 | 4.43739 | 7.43004 | 3.3662 | 51.66 | 0.03983 | 0.02303 |
| | Cl⁻ ($CaCl_2$) | 0.1003 | 4.15173 | 7.28211 | 3.53722 | 48.60 | 0.03936 | 0.02293 |
| | Cl⁻ ($MgCl_2$) | 0.10882 | 4.16289 | 11.54029 | 4.06613 | 56.06 | 0.04755 | 0.01446 |
| q = –84 | Cl⁻ (KCl) | 0.1017 | 4.40215 | 6.96218 | 3.33738 | 50.27 | 0.03897 | 0.02412 |
| | Cl⁻ (NaCl) | 0.09927 | 4.15825 | 6.47393 | 3.36329 | 46.74 | 0.03708 | 0.02583 |
| | Cl⁻ ($CaCl_2$) | 0.10017 | 4.19963 | 7.59406 | 3.56701 | 49.72 | 0.04017 | 0.02205 |
| | Cl⁻ ($MgCl_2$) | 0.10552 | 3.97389 | 13.03641 | 4.32569 | 56.16 | 0.04992 | 0.01286 |

**Table S.35.** Same as Table S.32, but for the case of DND–OH.

| DND's Charge | Dissolved Anion | $\tau_{in}^{dip}$ / ps | $\tau_{c}^{dip}$ / ps | $\tau_{m}^{dip}$ / ps | $\tau_{corr}^{dip}$ / ps | $\theta_{tot}^{dip}$ / ° | $d_{c}^{dip}$ / ps$^{-1}$ | $d_{m}^{dip}$ / ps$^{-1}$ |
|---|---|---|---|---|---|---|---|---|
| q = 0 | $Cl^-$ (KCl) | 0.10273 | 4.35392 | 6.85163 | 3.34214 | 49.60 | 0.0385 | 0.02448 |
| | $Cl^-$ (NaCl) | 0.10215 | 4.33332 | 6.92401 | 3.36584 | 49.41 | 0.03844 | 0.02433 |
| | $Cl^-$ ($CaCl_2$) | 0.10169 | 4.28269 | 7.74529 | 3.54251 | 50.57 | 0.04025 | 0.02183 |
| | $Cl^-$ ($MgCl_2$) | 0.08681 | 4.10702 | 10.85455 | 3.93843 | 55.45 | 0.04752 | 0.01557 |

## S.2.5.2. Water's OH reorientation parameters

**Table S.36.** The EWIC parameters for the OH reorientations of water in the first hydration shell of $Cl^-$ anions of different salts that are dissolved in the aqueous solutions of DND–H with various surface chemistries.

| DND's Charge | Dissolved Anion | $\tau_{in}^{oh}$ / ps | $\tau_{c}^{oh}$ / ps | $\tau_{m}^{oh}$ / ps | $\tau_{corr}^{oh}$ / ps | $\theta_{tot}^{oh}$ / ° | $d_{c}^{oh}$ / ps$^{-1}$ | $d_{m}^{oh}$ / ps$^{-1}$ |
|---|---|---|---|---|---|---|---|---|
| q = 0 | $Cl^-$ (KCl) | 0.07496 | 2.19878 | 5.25782 | 3.9301 | 27.25958 | 0.03213 | 0.03186 |
| | $Cl^-$ (NaCl) | 0.06072 | 2.42466 | 5.33355 | 3.95745 | 27.97642 | 0.03627 | 0.03151 |
| | $Cl^-$ ($CaCl_2$) | 0.07567 | 2.74527 | 5.1646 | 3.87307 | 27.64849 | 0.02497 | 0.03234 |
| | $Cl^-$ ($MgCl_2$) | 0.06571 | 3.102 | 5.49449 | 3.97053 | 29.88851 | 0.02774 | 0.03066 |
| q = +28 | $Cl^-$ (KCl) | 0.05929 | 1.8652 | 5.30959 | 4.00335 | 26.50711 | 0.03509 | 0.03149 |
| | $Cl^-$ (NaCl) | 0.08124 | 2.82702 | 5.50849 | 4.0569 | 28.5682 | 0.02438 | 0.03029 |
| | $Cl^-$ ($CaCl_2$) | 0.07408 | 2.51962 | 5.29185 | 3.95694 | 27.55377 | 0.02727 | 0.03159 |
| | $Cl^-$ ($MgCl_2$) | 0.06963 | 4.47112 | 6.22991 | 4.14595 | 35.78006 | 0.0232 | 0.02691 |
| q = +56 | $Cl^-$ (KCl) | 0.07348 | 4.65896 | 7.40723 | 4.57301 | 38.84231 | 0.02468 | 0.02264 |
| | $Cl^-$ (NaCl) | 0.0907 | 4.59397 | 7.10931 | 4.5421 | 37.28642 | 0.02348 | 0.02349 |
| | $Cl^-$ ($CaCl_2$) | 0.06931 | 4.62422 | 6.6516 | 4.33672 | 36.69061 | 0.02263 | 0.02511 |
| | $Cl^-$ ($MgCl_2$) | 0.07034 | 4.79916 | 7.34505 | 4.50463 | 39.44549 | 0.02457 | 0.02283 |
| q = +84 | $Cl^-$ (KCl) | 0.09219 | 4.86263 | 8.50258 | 5.0041 | 40.57234 | 0.02541 | 0.01966 |
| | $Cl^-$ (NaCl) | 0.09135 | 4.72841 | 8.37236 | 4.99535 | 39.80806 | 0.02532 | 0.01999 |
| | $Cl^-$ ($CaCl_2$) | 0.08892 | 5.2318 | 7.5464 | 4.55496 | 40.76633 | 0.0238 | 0.02213 |
| | $Cl^-$ ($MgCl_2$) | 0.09239 | 5.47368 | 8.63565 | 4.72704 | 44.61895 | 0.02612 | 0.01951 |

**Table S.37.** Same as Table S.36, but for the case of DND–$NH_2$.

| DND's Charge | Dissolved Anion | $\tau_{in}^{oh}$ / ps | $\tau_{c}^{oh}$ / ps | $\tau_{m}^{oh}$ / ps | $\tau_{corr}^{oh}$ / ps | $\theta_{tot}^{oh}$ / ° | $d_{c}^{oh}$ / $ps^{-1}$ | $d_{m}^{oh}$ / $ps^{-1}$ |
|---|---|---|---|---|---|---|---|---|
| q = 0 | $Cl^-$ (KCl) | 0.08089 | 3.01486 | 5.43968 | 3.97258 | 29.28 | 0.02407 | 0.03075 |
| | $Cl^-$ (NaCl) | 0.06551 | 2.56759 | 5.33065 | 3.98216 | 27.67 | 0.02659 | 0.03127 |
| | $Cl^-$ ($CaCl_2$) | 0.06293 | 2.62912 | 5.17598 | 3.90302 | 27.25 | 0.0284 | 0.03223 |
| | $Cl^-$ ($MgCl_2$) | 0.08721 | 4.96601 | 6.27592 | 4.1197 | 37.13 | 0.02175 | 0.02666 |
| q = +28 | $Cl^-$ (KCl) | 0.08155 | 2.89874 | 5.75035 | 4.16029 | 29.51 | 0.02607 | 0.02909 |
| | $Cl^-$ (NaCl) | 0.06477 | 2.73257 | 5.65733 | 4.13556 | 28.86 | 0.02551 | 0.02948 |
| | $Cl^-$ ($CaCl_2$) | 0.082 | 3.98649 | 5.74785 | 4.06664 | 32.00 | 0.02152 | 0.02904 |
| | $Cl^-$ ($MgCl_2$) | 0.06839 | 4.10656 | 6.06432 | 4.1617 | 33.70 | 0.02228 | 0.0275 |
| q = +56 | $Cl^-$ (KCl) | 0.07537 | 4.47183 | 7.35601 | 4.55179 | 38.42 | 0.02574 | 0.02283 |
| | $Cl^-$ (NaCl) | 0.09306 | 5.11155 | 7.69901 | 4.54368 | 41.25 | 0.02476 | 0.02196 |
| | $Cl^-$ ($CaCl_2$) | 0.0871 | 5.21988 | 6.61923 | 4.23616 | 38.53 | 0.02177 | 0.02524 |
| | $Cl^-$ ($MgCl_2$) | 0.09204 | 5.66075 | 7.71749 | 4.44123 | 43.80 | 0.02465 | 0.02162 |
| q = +84 | $Cl^-$ (KCl) | 0.10006 | 5.0739 | 13.68711 | 5.86106 | 49.85 | 0.0334 | 0.01229 |
| | $Cl^-$ (NaCl) | 0.09757 | 4.98022 | 14.84992 | 6.10119 | 50.26 | 0.03437 | 0.01145 |
| | $Cl^-$ ($CaCl_2$) | 0.07504 | 5.16016 | 10.24247 | 4.91232 | 48.20 | 0.03111 | 0.01659 |
| | $Cl^-$ ($MgCl_2$) | 0.09363 | 5.19919 | 11.87598 | 5.23552 | 50.18 | 0.03281 | 0.01426 |

**Table S.38.** Same as Table S.36, but for the case of DND–COOH.

| DND's Charge | Dissolved Anion | $\tau_{in}^{oh}$ / ps | $\tau_{c}^{oh}$ / ps | $\tau_{m}^{oh}$ / ps | $\tau_{corr}^{oh}$ / ps | $\theta_{tot}^{oh}$ / ° | $d_{c}^{oh}$ / $ps^{-1}$ | $d_{m}^{oh}$ / $ps^{-1}$ |
|---|---|---|---|---|---|---|---|---|
| q = 0 | $Cl^-$ (KCl) | 0.08017 | 3.17152 | 5.46387 | 3.9653 | 29.86 | 0.02437 | 0.03059 |
| | $Cl^-$ (NaCl) | 0.06838 | 3.11458 | 5.45194 | 3.98589 | 29.32 | 0.02458 | 0.03063 |
| | $Cl^-$ ($CaCl_2$) | 0.07746 | 2.75905 | 5.19926 | 3.90777 | 27.47 | 0.02364 | 0.03208 |
| | $Cl^-$ ($MgCl_2$) | 0.08575 | 4.7025 | 6.05898 | 4.0806 | 35.31 | 0.02091 | 0.02768 |
| q = –28 | $Cl^-$ (KCl) | 0.08465 | 3.93567 | 5.57838 | 3.9706 | 31.85 | 0.02462 | 0.03006 |
| | $Cl^-$ (NaCl) | 0.08114 | 3.50291 | 5.68098 | 4.01304 | 31.68 | 0.02602 | 0.02951 |
| | $Cl^-$ ($CaCl_2$) | 0.07164 | 2.1316 | 5.03584 | 3.87014 | 25.75 | 0.03167 | 0.03312 |
| | $Cl^-$ ($MgCl_2$) | 0.08186 | 3.89672 | 5.9299 | 4.09012 | 33.27 | 0.02299 | 0.02822 |
| q = –56 | $Cl^-$ (KCl) | 0.07893 | 3.11591 | 5.38601 | 3.95653 | 29.23 | 0.02696 | 0.03103 |
| | $Cl^-$ (NaCl) | 0.09174 | 5.08129 | 6.40576 | 3.98448 | 40.01 | 0.02332 | 0.02689 |
| | $Cl^-$ ($CaCl_2$) | 0.07368 | 2.18292 | 5.10083 | 3.87894 | 26.33 | 0.02762 | 0.0327 |
| | $Cl^-$ ($MgCl_2$) | 0.07009 | 4.24562 | 5.94324 | 4.08192 | 33.93 | 0.02235 | 0.02812 |
| q = –84 | $Cl^-$ (KCl) | 0.08434 | 3.38336 | 5.45375 | 3.97924 | 29.69 | 0.02143 | 0.0306 |
| | $Cl^-$ (NaCl) | 0.07915 | 2.51967 | 5.36294 | 4.00315 | 27.62 | 0.02594 | 0.03112 |
| | $Cl^-$ ($CaCl_2$) | 0.07777 | 2.93266 | 5.36266 | 3.95985 | 28.59 | 0.02331 | 0.03112 |
| | $Cl^-$ ($MgCl_2$) | 0.08632 | 4.77948 | 6.22799 | 4.12085 | 36.34 | 0.02155 | 0.02692 |

**Table S.39.** Same as Table S.36, but for the case of DND–OH.

| DND's Charge | Dissolved Anion | $\tau_{in}^{oh}$ / ps | $\tau_{c}^{oh}$ / ps | $\tau_{m}^{oh}$ / ps | $\tau_{corr}^{oh}$ / ps | $\theta_{tot}^{oh}$ / ° | $d_{c}^{oh}$ / ps$^{-1}$ | $d_{m}^{oh}$ / ps$^{-1}$ |
|---|---|---|---|---|---|---|---|---|
| q = 0 | $Cl^-$ (KCl) | 0.07422 | 1.89694 | 5.09847 | 3.92837 | 25.41 | 0.03056 | 0.03271 |
| | $Cl^-$ (NaCl) | 0.06327 | 2.49872 | 5.32979 | 3.97883 | 27.66 | 0.03382 | 0.03141 |
| | $Cl^-$ ($CaCl_2$) | 0.06349 | 3.11982 | 5.41132 | 3.92411 | 30.27 | 0.03001 | 0.03107 |
| | $Cl^-$ ($MgCl_2$) | 0.08629 | 4.77373 | 6.04898 | 4.07268 | 35.59 | 0.02076 | 0.02767 |